\documentstyle[pra,aps,graphicx,preprint]{revtex}

\begin{document}
\bibliographystyle{prsty}

\draft
\title{
Coordinate-space approach to the bound-electron self-energy: \\
       Self-Energy screening calculation       }
\author{P. Indelicato}
\address{
Laboratoire Kastler-Brossel,
 Unit\'e Mixte de Recherche du CNRS n$^{\circ}$ C8552,\\
 \'Ecole Normale Sup\'erieure et Universit\'e Pierre et Marie 
 Curie, Case 74, 4 place Jussieu,
 F-75252 Paris CEDEX 05, France  }
\author{Peter J. Mohr}
\address{
  National Institute of Standards and Technology,
  Gaithersburg, Maryland 20899-8401
 }
\date{ Time-stamp: <Tuesday, October 17, 2000, 10:21:29 dft paul>  } 
\maketitle 
\begin{abstract}
The self-energy screening correction is evaluated in a model in which
the effect of the screening electron is represented as a first-order
perturbation of the self energy by an effective potential.  The
effective potential is the Coulomb potential of the
spherically averaged charge density of the screening electron.  We
evaluate the energy shift due to a $1s_{1/2}$, $2s_{1/2}$, $2p_{1/2}$,
or $2p_{3/2}$ electron screening a $1s_{1/2}$, $2s_{1/2}$, $2p_{1/2}$,
or $2p_{3/2}$ electron, for nuclear charge Z in the range $5 \le Z\le
92$.  A detailed comparison with other calculations is made.
\end{abstract}
\pacs{31.30.Jv}
\narrowtext


\section{introduction}
\label{sec:intro}

\begin{figure}
\centering
\includegraphics{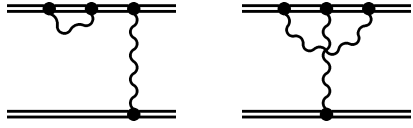}
\caption{Feynman diagrams for radiative corrections to the
electron-electron interaction. (A) represents the wave function
correction and (B) is the vertex correction.
\label{fig:diag}}
\end{figure}

\begin{figure}
\centering
\includegraphics{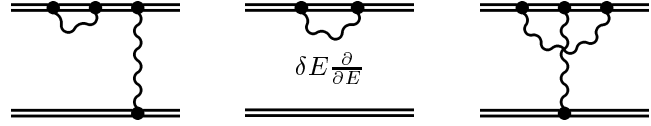}
\caption{Feynman diagrams for radiative corrections to the
electron-electron interaction after extraction of reducible (A) and
irreducible (A') parts for the wave function correction (A) of
Fig. \protect \ref{fig:diag}.
\label{fig:diagred}}
\end{figure}

The self-energy correction to the electron-electron interaction is one
of the many contributions of order $\alpha^2$ to an atomic binding
energy.  These corrections, shown as Feynman diagrams
in Fig.~\ref{fig:diag}, are often called self-energy screening corrections, 
and for inner shells they are the
largest of all fourth-order radiative corrections.  They give rise to
three terms,  which are represented in Fig.~\ref{fig:diagred},
when one distinguishes between the reducible and
irreducible part of the diagram in Fig.~\ref{fig:diag} (A).

\begin{figure}
\centering
\includegraphics{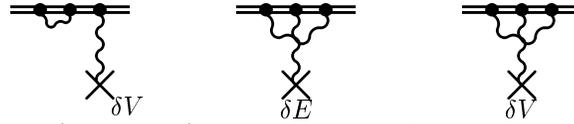}
 \caption{Feynman diagrams for the self-energy perturbed by an external potential.
   In the case of the spherically-averaged screening potential used in this work, 
   the diagram in the left originates in the diagram (A) in Fig.
   \protect \ref{fig:diagred}, and represents the wave function
   correction, while the diagram in the center comes from diagram (A')
   in Fig. \protect \ref{fig:diagred}, using the (symbolic) relation
   $\frac{\partial}{\partial E} G(E)=G(E)\cdot G(E)$. The diagram on
   the right comes from the vertex correction.
\label{fig:diag_s}}
\end{figure}

A first attempt to evaluate the contribution of such diagrams from
bound state quantum electrodynamics (BSQED) was made in 1991
\cite{iam91} in an approximation in which the electrons not associated
with the self-energy loop were represented as a perturbing potential
(Fig.~\ref{fig:diag_s}).  The potential was obtained by taking a
spherical average of the electron wave function, and calculating the
potential associated with the resulting charge density.  More
recently, direct evaluations of the diagrams of Fig.~\ref{fig:diag}
for the ground state of two-electron ions have been made
\cite{yas95,pssl96,yas97,yab98,yabss98,yab99}, and the method of
\cite{iam91} was used to provide the self-energy correction to several
interactions \cite{bcs97}.  In this paper we report on a complete
calculation of the self-energy screening in the approximation of
\cite{iam91}, for all combinations of pairs of the states $1s_{1/2}$,
$2s_{1/2}$, $2p_{1/2}$, or $2p_{3/2}$ in the range $5\leq Z \leq 92$.

Terms corresponding to the diagram in Fig.~\ref{fig:diag_s}(A) are
obtained by varying the self-energy expression with respect to the
external potential of the 
bound-state wave function, leading to the introduction of the
first-order correction to the wave function in the potential $\delta
V$.  Terms corresponding to the diagram in
Fig.~\ref{fig:diag_s}(B) are obtained by varying the expression for
the self-energy with respect to the external potential of the 
Green's
function and while those in Fig.~\ref{fig:diag_s}(A$^\prime$) correspond to
variation with respect to the energy of the bound state.

The expression for the self energy in a large class of potentials 
${\cal V}(\bbox{x})$ can be written as
the sum $ {\cal E}_{\rm SE} = {\cal E}_{\rm L} + {\cal E}_{\rm H}$ 
of a low-energy part $ {\cal E}_{\rm L}$ and a high-energy part 
$ {\cal E}_{\rm H}$ given (in units in which $\hbar=c=m_{\rm e} = 1)$
by \cite{moh74a}
\widetext
\begin{equation} 
 {\cal E}_{\rm L}=\frac{\alpha} {\pi} {\cal E}_n - \frac{\alpha} {\pi}
{\rm P}\!\int_0^{{\cal E}_n}
d z 
\int d\bbox x_2 \int  d\bbox x_1 \, \varphi_n^\dagger (\bbox x_2) \alpha^l
{\cal G}(\bbox x_2,\bbox x_1, z) \alpha^m \varphi_n(\bbox x_1) (\delta_{lm}\bbox
\nabla_2\cdot \bbox \nabla_1 - \nabla_2^l \nabla_1^m) \frac{\sin 
[({\cal E}_n-z)x_{21}] }{({\cal E}_n-z)^2 x_{21}}
\label{eq:low_e} \end{equation} 
\noindent and
\begin{equation} 
\  {\cal E}_{\rm H}=\frac{\alpha}{ 2\pi {\rm i}} \int_{\rm C_H} d z\ \int d\bbox x_2 \int
d\bbox x_1 \, \varphi_n^\dagger (\bbox x_2) \alpha_\mu 
{\cal G}(\bbox x_2,\bbox x_1, z)
 \alpha^\mu \varphi_n(\bbox x_1) \frac{e^{-bx_{21}}}{ x_{21}} -\delta m\int 
  d\bbox x\ \varphi_n^\dagger (\bbox x) \beta \varphi_n(\bbox x) \ , 
\label{eq:high_e}
\end{equation} 
\narrowtext \noindent where $b=-i\left[({\cal E}_n-z)^2 
+i\delta\right]^{1/2}, {\rm Re}(b)>0$, and $\bbox x_{21} = \bbox x_2 - 
\bbox x_1$.  In these expressions, $\varphi_n$ and ${\cal E}_n$ are the 
eigenfunction and eigenvalue of the Dirac equation for the bound state 
$n$, and ${\cal G}$ is the Green's function for the Dirac equation 
corresponding to the operator ${\cal G} = ({\cal H}-z)^{-1}$, where 
${\cal H}=\bbox 
\alpha\cdot\bbox p + {\cal V} + \beta$ is the Dirac Hamiltonian.  The indices 
$l$ and $m$ are summed from 1 to 3, and the index $\mu$ is summed from 
0 to 3.  The contour ${\rm C_H}$ extends from $-i\infty$ to $0-i\epsilon$ 
and from $0+i\epsilon$ to $+i\infty$, with the appropriate branch of 
$b$ chosen in each case. For the present calculation, we assume that 
the potential  ${\cal V}(\bbox{x})$ is close to a pure Coulomb 
potential, except for a small correction $\delta V(\bbox{x})$, which 
is not necessarily spherically symmetric. Indeed some applications 
of this method have been made with non-spherically-symmetric perturbations 
\cite{bcs97}.
We obtain the screening correction to the self-energy
by making the replacements
\begin{mathletters}
\begin{eqnarray}
    {\cal V}(\bbox{x}) &\to & V(x) + \delta V(\bbox{x}) \ , 
    \label{eq:repv}\\
\varphi_n(\bbox x) &\to & \phi_n(\bbox x) + \delta \phi_n(\bbox x) \ , 
\label{eq:repwf} \\
{\cal G}({\bbox x}_2,{\bbox x}_1,z)& \to& G({\bbox x}_2,{\bbox x}_1,z)+ \delta 
G({\bbox x}_2,{\bbox x}_1,z) \label{eq:repgr}  \\
{\cal E}_n& \to &E_n +  \delta E_n \label{eq:repen}
\end{eqnarray}
\end{mathletters}

\noindent in Eqs.~(\ref{eq:low_e}) and (\ref{eq:high_e}) and 
retaining only the
first-order correction terms.  In the Eqs.~(\ref{eq:low_e}) and
(\ref{eq:high_e}), and (\ref{eq:repv}) to (\ref{eq:repen}) above, we
use the symbols ${\cal G}$, ${\cal E}_{n}$, $\varphi_{n}$ for the
{\it exact} quantities in the potential ${\cal V}(\bbox{x})$, while
the symbols ${G}$, ${E}_{n}$, $\phi_{n}$ represent 
the corresponding {\it exact}
quantities in a pure Coulomb potential ${V}({x})$.
The same conventions are employed throughout the paper.
We denote the operations on the unperturbed
self energy that lead to these three corrections by $\delta_{\phi}$,
$\delta_{\rm G}$, and $\delta_{\rm E}$, respectively.  
In particular, we employ the notation
\begin{eqnarray}
\delta_\phi &=& \delta \phi_n \frac{\partial}{\partial \varphi_n} \\
\delta_G &=& \delta G \frac{\partial}{\partial {\cal G}} \\
\delta_{\rm E} &=& \delta E_n \frac{\partial}{\partial {\cal E}_n} \ ,
\end{eqnarray}
and the total correction is the sum
\begin{eqnarray} 
\delta \phi_n \frac{\partial}{\partial \varphi_n}
+ \delta G \frac{\partial}{\partial {\cal G}}
+ \delta E_n \frac{\partial}{\partial {\cal E}_n}
\label{eq:parts}
\end{eqnarray}
in which the partial differentiation symbol denotes 
formal differentiation with respect to the indicated variable,
with the result evaluated with the unperturbed functions.

In Sec.  \ref{sec:pert} we write expressions for
the first-order perturbation
corrections to the energy, the wave function and the Green's function.
In Secs.  \ref{sec:lowpar}, \ref{sec:high_e}, and \ref{sec:an_sub} we
derive the expressions for the various contributions to the
screened self-energy corresponding to the three diagrams of 
Fig.~\ref{fig:diag_s}.  In a series of three earlier papers
\cite{iam92,iam95,iam98}, we have derived and tested a method
of analytically isolating divergent contributions to the 
self-energy diagram in coordinate
space.  In Secs.  \ref{sec:high_e} and \ref{sec:an_sub} we derive from
this earlier work the generalizations of the 
analytic subtraction terms which are necessary
to make all contributions to  the self-energy screening finite. The 
numerical results are presented
in Sec.~\ref{sec:results}, and Sec.~\ref{sec:concl} is the conclusion.

\section{Perturbation expansion}
\label{sec:pert}
Replacing the potential ${\cal V}({\bbox x})$ 
by $V(x)+\delta V(x)$, where 
$\delta V(x)$ is spherically symmetric, changes the wave functions, 
the energy, and the Green's function, which appear in 
Eqs.~(\ref{eq:low_e}) and (\ref{eq:high_e}).  From standard perturbation
theory the first-order energy correction is
\begin{eqnarray}
\delta E_n &=& \int d \bbox x \, \phi^\dagger_n(\bbox x) \,\delta V(x)
\phi_n(\bbox x) 
\nonumber\\
&=& \int_0^\infty dx \, x^2 \, \sum_{i = 1}^2 f_{n,i}(x)\,\delta V(x)
f_{n,i}(x) \ ,
\label{eq:foen}
\end{eqnarray}
where the radial wave function $f_{n,i}$ is defined by writing
\begin{equation} 
\phi_n(\bbox x) = \left[\begin{array}{c}
f_{n,1}(x)\chi_\kappa^\mu(\hat x) \\ 
 {\rm i}f_{n,2}(x)\chi_{-\kappa}^\mu(\hat x) \end{array}\right] \ , 
\label{eq:dwf}
\end{equation}
and where $\chi_\kappa^\mu(\hat x)$ is the Dirac angular momentum
eigenfunction.

The first-order 
correction to the wave function is given with the aid of the reduced 
Green's function $G^{R}(\bbox x_2,\bbox x_1,E_n)$,
defined by
(see for example \cite{hyl84})
\begin{eqnarray}
&&G^{R}(\bbox x_2,\bbox x_1,E_n)
=\sum_{\stackrel{\scriptstyle m}{\scriptstyle E_m \ne E_n }}
\frac{\phi_{m}(\bbox x_2)\phi_{m}^{\dagger}(\bbox x_1)
                            }{ E_m - E_n}
\nonumber\\
&&\qquad =\lim_{z \rightarrow E_n}
\left[G(\bbox x_2,\bbox x_1,z) - 
\frac{\phi_{n}(\bbox x_2)\phi_{n}^{\dagger}(\bbox x_1)
                            }{ E_n - z}\right] \ ,
\label{eq:redgreen}
\end{eqnarray}
as
\begin{eqnarray}
\delta\phi_n(\bbox{x}_2) =
-\int d \bbox{x}_1 G^{\rm R}(\bbox {x}_2, \bbox{x}_1, E_n)
\,\delta V(x_1)\phi_n(\bbox x_1) \ .
\label{eq:wfcorr}
\end{eqnarray}
For a spherically symmetric potential we have
\begin{eqnarray}
&&\delta f_{n,i}(x_2)
\nonumber\\
&&=-\int_0^{\infty} d x_1 x_1^2 \sum_{j=1}^{2}
G_{\kappa}^{R,ij}(x_2,x_1,E_n)\,\delta V(x_1) f_{n,j}(x_1) \ .
\label{eq:radwfcorr}
\end{eqnarray}
In Eq.~(\ref{eq:radwfcorr}), the components of the radial reduced
Green's function $G_{\kappa}^{R,ij}(x_2,x_1,E_n)$ are defined in
analogy with the components of the full Green's function
$G_{\kappa}^{ij}(x_2,x_1,z)$ as given in Ref.~\cite{moh74a}, 
Eq.~(A.14).

To evaluate the first-order correction to the Green's function we use 
the well-known expansion
\begin{eqnarray}
{\cal G}(z) &=& \frac{1}{ H + \delta V - z} \nonumber \\
     &=& \frac{1}{ H - z} - \frac{1}{ H - z} \,\delta V \frac{1}{ H - z}
     \nonumber \\&&+ \frac{1}{ H - z} \,\delta V \frac{1}{ H - z}
      \,\delta V \frac{1}{ H - z}
      + \cdots \ ,
\label{eq:fgreenexp}
\end{eqnarray}        
and the term of first order in $\delta V$ is
\begin{eqnarray}
\delta G(z) &=& - \frac{1}{H - z} \,\delta V \frac{1}{ H - z} \ ,
\label{eq:fdv}
\end{eqnarray}
which has second-order poles at the eigenvalues of the Dirac equation.  
In coordinate space, the first-order correction is
\begin{eqnarray}
\lefteqn{\delta G({\bbox x}_2,{\bbox x}_1,z) =}&& \nonumber \\ 
&& \qquad - \int d{\bbox x}_3 \,G({\bbox x}_2,{\bbox x}_3,z)
   \, \delta V(x_3)G({\bbox x}_3,{\bbox x}_1,z) \ ,
\label{eq:fgreencorr}
\end{eqnarray}
and for $\delta V(x_3)$ spherically symmetric, we have
\begin{eqnarray}
\lefteqn{\delta G_\kappa^{ij}(x_2,x_1,z)
=- \int_0^\infty dx_3 \, x_3^2 }&& \nonumber \\&& 
\qquad\times \sum_{k=1}^2 G_\kappa^{ik}(x_2,x_3,z) 
\,\delta V(x_3) G_\kappa^{kj}(x_3,x_1,z) \ .
\label{eq:fgreencors}
\end{eqnarray}


\section{Low-Energy Part}
\label{sec:lowpar}
The low-energy part, for an arbitrary external spherically symmetric
potential, when
integrated over the spherical angles of the vectors ${\bbox x}_2$ and
${\bbox x}_1$, yields
\begin{eqnarray}
  {\cal E}_{\rm L} =\frac{\alpha}{\pi}{\cal E}_n+\frac{\alpha}{\pi}P\int_0^{{\cal E}_n}dz
   \,  {\cal U}(z)
\label{eq:lowe_ang}
\end{eqnarray}
with
\begin{eqnarray}
\lefteqn{{\cal U}(z) =-\int_0^\infty dx_2 x_2^2 \int_0^\infty dx_1x_1^2} &&\nonumber \\
&&\times \sum_\kappa\sum_{i,j=1}^2{\cal F}_{n,\bar \imath}(x_2) {\cal G}_\kappa^{ij}
(x_2,x_1,z) {\cal F}_{n,\bar \jmath}(x_1) {\cal A}_\kappa^{ij}(x_2,x_1,z), 
\label{eq:2}
\end{eqnarray}
where the summation over $\kappa$ runs over all nonzero integers, and
where $\bar \imath = 3 - i$ and $\bar \jmath = 3 - j$.

We are concerned with the first-order perturbation in this expression 
that arises from variation of the external potential.  The self energy 
depends on the potential through three quantities that appear in 
Eq.~(\ref{eq:lowe_ang}) and (\ref{eq:2}), the wave function 
${\cal F}_{n,i}(x)$, 
the energy eigenvalue ${\cal E}_n$, and the 
Green's function ${\cal G}_\kappa^{ij} (x_2,x_1,z)$.  The three 
corrections are denoted by $\delta_{\rm \phi} \, {\cal E}_{\rm L}$, 
$\delta_{\rm E} \,{\cal 
E}_{\rm L}$, and $\delta_{\rm G} \, {\cal E}_{\rm L}$, 
respectively, with the total
\begin{eqnarray}
\delta E_{\rm L} = \delta_{\rm \phi} \, {\cal E}_{\rm L}
+ \delta_{\rm E} \,{\cal E}_{\rm L}
+ \delta_{\rm G} \, {\cal E}_{\rm L} \ .
\end{eqnarray}


\subsection{
Lower-order terms}
\label{sec:loword}
The expression (\ref{eq:low_e}) contains spurious parts of lower order
in $Z\alpha$ than the complete result.  The physically significant part
is isolated in a function
${\cal F}_{\rm L}(Z\alpha)$ defined by~\cite{iam92}
\begin{eqnarray}
 {\cal E}_{\rm L} &=& \frac{\alpha}{\pi}\bigg[\frac{5}{ 6} {\cal E}_n 
+ \frac{2}{ 3} \left<\varphi_n|\beta|\varphi_n\right>
     +\frac{7}{ 6}\left<\varphi_n|{\cal V}|\varphi_n\right>
\nonumber\\&&\qquad \qquad
 + \frac{(Z\alpha)^4}{ n^3}{\cal F}_{\rm L}(Z\alpha)\bigg]
\label{eq:flsh}
\end{eqnarray}
Here the linear perturbation of this function with respect to variation of the
external potential is of interest.  A corresponding function 
$\delta F_{\rm L}(Z\alpha)$ is defined by
\begin{eqnarray}
\delta  {E}_{\rm L} &=& \frac{\alpha}{\pi}\bigg[2\,\delta E_n 
     + \frac{4}{ 3} \left<\phi_n|\beta|\delta\phi_n\right>
     +\frac{7}{ 3}\left<\phi_n|V|\delta\phi_n\right> 
\nonumber\\&&\qquad \qquad
     + \frac{\alpha(Z\alpha)^3}{ n^3}\delta F_{\rm L}(Z\alpha)\bigg]
\label{eq:flshp}
\end{eqnarray}
where the fact that 
\begin{eqnarray}
\delta\left<\phi_n|V|\phi_n\right> &=& 
\left<\phi_n|\delta V|\phi_n\right> + 2 \left<\phi_n|V|\delta\phi_n\right>
\nonumber\\
&=& \delta E_n + 2 \left<\phi_n|V|\delta\phi_n\right>
\end{eqnarray}
has been taken into account.

\subsection{
Low-order matrix elements}
\label{sec:matele}

The lower-order expectation values involving the first-order
correction to the wave function $\delta \phi_n$ can be evaluated by
direct numerical integration. However, to get an independent check of
the precision of the calculation, particularly when strong
cancellation occurs, we derive a number of useful expressions in which
$\delta \phi_n$ does not occur.  The energy perturbation is given by the
conventional expression
\begin{eqnarray}
\delta E_n &=& \left<\phi_n|\delta V|\phi_n\right>
\nonumber\\
&=&
\int_0^\infty dx\ x^2\left[f_1^2(x) + f_2^2(x)\right]\delta V(x) \ .
\end{eqnarray}

The wave-function perturbation terms are
\begin{eqnarray}
\left<\phi_n|\beta|\delta \phi_n\right> &=& 
\int  d\bbox{x}_2 \int d\bbox{x}_1 \, \phi_n^\dagger (\bbox{x}_2) \,
\beta\nonumber\\ &&
\times \sum_{E_i \ne E_n}\frac{\phi_i (\bbox{x}_2)\phi_i^\dagger (\bbox{x}_1)
}{ E_n - E_i} \, \delta V(x_1) \phi_n(\bbox{x}_1)
\label{eq:bpert}
\end{eqnarray}
and
\begin{eqnarray}
\left<\phi_n|V|\delta \phi_n\right> &=&
\int  d\bbox{x}_2 \int d\bbox{x}_1 \, \phi_n^\dagger (\bbox{x}_2) \,
V(x_2) \nonumber\\ &&
\times \sum_{E_i \ne E_n}\frac{\phi_i (\bbox{x}_2)\phi_i^\dagger (\bbox{x}_1)
}{ E_n - E_i} \, \delta V(x_1) \phi_n(\bbox{x}_1) \ .
\label{eq:vpert}
\end{eqnarray}   
Since we are considering here the case where the unperturbed potential 
is the Coulomb potential with known wave functions, we can simplify 
the calculation of these matrix elements.  In particular, we interpret 
the expressions in Eqs.~(\ref{eq:bpert}) and (\ref{eq:vpert}) as 
perturbations of the wave function on the left-hand side to give
\begin{eqnarray}
\left<\phi_n|\beta|\delta \phi_n\right> = 
\left<\delta_\beta \phi_n|\delta V|\phi_n\right> 
\end{eqnarray}
and
\begin{eqnarray}
\left<\phi_n|V|\delta \phi_n\right> = 
\left<\delta_V \phi_n|\delta V| \phi_n\right>  \ ,
\end{eqnarray}
where $\delta_\beta \phi_n$ and $\delta_V \phi_n$ are first-order
corrections to the wave function due to perturbations $\beta$
and $V$, respectively.  These are simply calculated as the 
coefficients of $\delta$ in the power series expansions 
in $\delta$ of the
wave functions for the appropriately modified Hamiltonians
\begin{eqnarray}
H_\beta &=& \bbox{\alpha}\cdot\bbox{p} + (1+\delta)\beta + V
\nonumber\\
        &=& \bbox{\alpha}\cdot\bbox{p} + (1+\delta)\beta m + V
\label{eq:hamb}
\end{eqnarray}
and
\begin{eqnarray}
H_V = \bbox{\alpha}\cdot\bbox{p} + \beta + (1+\delta)V \ .
\label{eq:hamv}
\end{eqnarray}
In (\ref{eq:hamb}) the second line restores the mass dependence of the 
Hamiltonian in order to exhibit the dependence of the modified Coulomb 
wave functions on $\delta$.  

The wave function correction 
$\delta_\beta \phi_n$ is obtained by replacing $m$ by $(1+\delta)m$ in 
$\phi_n$ and calculating the coefficient of the term linear in 
$\delta$, which is equivalent to writing
\begin{eqnarray}
\delta_\beta \phi_n = m\frac{\partial}{\partial m}\phi_n
\end{eqnarray} 
and leads to 
\begin{eqnarray}
\left<\phi_n|\beta|\delta \phi_n\right> = 
\frac{m}{2}\frac{\partial}{\partial m}\left<\phi_n|\delta V|\phi_n\right> \ ,
\label{eq:bterm}
\end{eqnarray}
where the derivative acts only on the wave function.
A further simplification is possible based on the mass dependence of the
wave function in (\ref{eq:bterm}).  If $x$ is replaced by $x/m$ in the 
second line of (\ref{eq:hamb}), then the mass factors out of the
Hamiltonian, and the wave function is independent of $m$.  As a result, 
we have
\begin{eqnarray}
\left<\phi_n|\beta|\delta \phi_n\right> &=&
\frac{m}{2}\frac{\partial}{\partial m}
\int_0^\infty dx\ x^2\left[f_1^2(x) + f_2^2(x)\right]\delta V(x)
\nonumber\\ &=&
\frac{1}{m^3}\int_0^\infty dx\ x^2
\nonumber \\ &&\times
\left[f_1^2(x/m) + f_2^2(x/m)\right]
\frac{m}{2}\frac{\partial}{\partial m} \delta V(x/m)
\nonumber\\ &=&
-\frac{1}{2}\int_0^\infty dx\ x^3
\left[f_1^2(x) + f_2^2(x)\right]
\frac{\partial}{\partial x} \delta V(x)
\nonumber\\ &=&-{1\over2}\left<\phi_n|x\,(\delta V)^\prime |\phi_n\right> \ ,
\label{eq:memass}
\end{eqnarray}
where the convention $m = 1$ has been restored in the last line, and
$(\delta V)^\prime$ represents the operator corresponding to
$\frac{\partial}{\partial x} \delta V(x)$.  

The correction $\delta_V
\phi_n$ is obtained by replacing $Z$ by $(1+\delta)Z$ in $\phi_n$ and
calculating the coefficient of the term linear in $\delta$ which is
equivalent to writing
\begin{eqnarray}
\delta_V \phi_n = Z\frac{\partial}{\partial Z}\phi_n
\end{eqnarray}
which leads to
\begin{eqnarray}
\left<\phi_n|V|\delta \phi_n\right> = 
\frac{Z}{2}\frac{\partial}{\partial Z}\left<\phi_n|\delta V|\phi_n\right>
\end{eqnarray}
where it is understood that the derivative acts only on the wave
function, or
\begin{eqnarray}
\left<\phi_n|V|\delta \phi_n\right> &=&
Z \int_0^\infty dx\ x^2
\nonumber\\ 
&& \times \left[f_1(x)\frac{\partial f_1(x)}{\partial Z} 
+ f_2(x)\frac{\partial f_2(x)}{\partial Z}\right]\delta V(x) \ .
\nonumber\\
\label{eq:vme}
\end{eqnarray}
The derivative of the potential in (\ref{eq:memass}) is calculated
analytically, and the derivatives in (\ref{eq:vme}) are calculated
numerically by evaluating the wave function with 32 figure precision
and using a symmetric derivative formula with $\delta(Z\alpha) =
10^{-12}$.

The matrix elements listed above are evaluated by Gaussian quadrature,
and the code was tested in the Coulomb case and compared to the
analytic results, as described in Appendix \ref{app:numtest}.

\subsection{
Low-Energy energy-level and wave function correction}
\label{sec:loenper}
The correction from the energy-level perturbation of the low-energy 
part of the self energy (\ref{eq:low_e}) is
\begin{eqnarray}
\delta_{\rm E}  {\cal E}_{\rm L} = \delta E_n 
\frac{\partial  {\cal E}_{\rm L}}{ \partial {\cal E}_n} \ ,
\end{eqnarray}
where
\widetext
\begin{eqnarray}
\frac{\partial  {\cal E}_{\rm L}}{ \partial {\cal E}_n}
&=& \frac{\alpha}{\pi}
 - \frac{\alpha}{\pi}\left[
\int d\bbox{x}_2 \int d\bbox{x}_1 \ \phi_n^\dagger (\bbox{x}_2) \alpha^l
G(\bbox{x}_2,\bbox{x}_1, z) \alpha^m \phi_n(\bbox{x}_1) 
(\delta_{lm}\bbox{
\nabla}_2\cdot \bbox{\nabla}_1 - \nabla_2^l \nabla_1^m) \frac{\sin
[(E_n-z)x_{21}] }{(E_n-z)^2 x_{21}} \right]_{z = E_n}
\nonumber\\ &&-
\frac{\alpha}{\pi}{\rm P}\!\int_0^{E_n}dz
\int d\bbox{x}_2 \int d\bbox{x}_1 \ \phi_n^\dagger (\bbox{x}_2) \alpha^l
G(\bbox{x}_2,\bbox{x}_1, z) \alpha^m \phi_n(\bbox{x}_1) (\delta_{lm}\bbox{
\nabla}_2\cdot \bbox{\nabla}_1 - \nabla_2^l \nabla_1^m) 
\frac{\partial}{\partial E_n}
\frac{\sin [(E_n-z)x_{21}] }{(E_n-z)^2 x_{21}} \ .
\label{eq:eder}
\end{eqnarray}
The second term on the right-hand side of (\ref{eq:eder}) 
makes no contribution because
\begin{eqnarray}
(\delta_{lm}\bbox{
\nabla}_2\cdot \bbox{\nabla}_1 - \nabla_2^l \nabla_1^m) \frac{\sin
[(E_n-z)x_{21}] }{(E_n-z)^2 x_{21}}
&=&\frac{2}{3}\delta_{lm}(E_n - z) + {\cal O}(E_n - z)^3 \ ,
\end{eqnarray}
and
\begin{eqnarray}
\int d\bbox{x}_2 \int d\bbox{x}_1 \
\phi_n^\dagger (\bbox{x}_2) \alpha^l
G(\bbox{x}_2,\bbox{x}_1, z) \alpha^m \phi_n(\bbox{x}_1)
&=&\int d\bbox{x}_2 \int d\bbox{x}_1 \
\phi_n^\dagger (\bbox{x}_2) \alpha^l
\sum_{E_j = E_n}\frac{\phi_j(\bbox{x}_2)\phi_j^\dagger (\bbox{x}_1)
}{ E_n - z}
\alpha^m \phi_n(\bbox{x}_1) + {\cal O}(1)
\nonumber\\
&=& {\cal O}(1) \ .
\label{eq:surfterm}
\end{eqnarray}
The first term on the right-hand side of Eq.~(\ref{eq:surfterm}) vanishes 
by virtue of the identity
\begin{eqnarray}
\int d\bbox{x} \ \phi_n^\dagger (\bbox{x}) \alpha^l
\phi_j(\bbox{x})
&=& {\rm i} \int d\bbox{x} \ \phi_n^\dagger (\bbox{x})[H,x^l]
\phi_j(\bbox{x}) 
= {\rm i}\,(E_n - E_j)\int d\bbox{x} \ \phi_n^\dagger (\bbox{x})x^l
\phi_j(\bbox{x}) \ .
\label{eq:hamilcom}
\end{eqnarray}
\narrowtext \noindent
The correction due to variation of the bound state eigenvalue is thus
\begin{eqnarray}
\delta_{\rm E}  {\cal E}_{\rm L} 
&=&  \frac{\alpha}{\pi}\delta E_n + 
\frac{\alpha}{\pi}P\int_0^{E_n}dz
   \, \delta_{\rm E} {\cal U}(z) \ ,
\end{eqnarray}
where
\begin{eqnarray}
\lefteqn{\delta_{\rm E} \, {\cal U}(z) 
= - \delta E_n\int_0^\infty dx_2 \, x_2^2 \int_0^\infty dx_1 \, x_1^2
\sum_\kappa \sum_{i,j=1}^2 }&&
 \nonumber \\ &&
f_{\bar \imath}(x_2) G_\kappa^{ij}
(x_2,x_1,z) f_{\bar \jmath}(x_1) \frac{\partial}{\partial E_n} A_\kappa^{ij}(x_2,x_1,z)  \ .
\label{eq:deang}
\end{eqnarray}

From Eq.~(\ref{eq:low_e}) one can also easily obtain the correction due to 
the variation of the bound state wave function:
\widetext
\begin{equation} 
\delta_{\phi} {\cal E}_{\rm L}
= - \frac{2\alpha} {\pi}{\rm P}\!\int_0^{E_n}
 d z 
\int d\bbox x_2 \int d\bbox x_1 \ \phi_n^\dagger (\bbox x_2) \alpha^l
G(\bbox x_2,\bbox x_1, z) \alpha^m \delta\phi_n(\bbox x_1) (\delta_{lm}\bbox
\nabla_2\cdot \bbox \nabla_1 - \nabla_2^l \nabla_1^m) \frac{\sin 
[(E_n-z)x_{21}] }{(E_n-z)^2 x_{21}} \ ,
\label{eq:low_edwf}
\end{equation}
\narrowtext \noindent
since the dependence on the bound-state wave function is explicit.

\subsection{Low-Energy Green's function correction}
\label{sec:lognper}
The correction due to the variation of the Green's function in
Eq.~(\ref{eq:lowe_ang}) is given by
\begin{eqnarray}
\delta_{\rm G}  {\cal E}_{\rm L} 
&=&  \frac{\alpha}{\pi} \,
{\rm Re} \int_{{\rm C}_+}dz\,
\delta_{\rm G} \, {\cal U}(z)
\label{eq:lowedgang}
\end{eqnarray}
where first-order change in ${\cal U}(z)$ in Eq.~(\ref{eq:2}), due to
variation of ${\cal G}$ is
\begin{eqnarray}
\lefteqn{\delta_{\rm G} \, {\cal U}(z) 
= - \int_0^\infty dx_2 \, x_2^2 \int_0^\infty dx_1 \, x_1^2}&&
\nonumber \\ && \times\sum_\kappa \sum_{i,j=1}^2 
f_{n,\bar \imath}(x_2) \, \delta G_\kappa^{ij}
(x_2,x_1,z) f_{n,\bar \jmath}(x_1) A_\kappa^{ij}(x_2,x_1,z) \ ,
\nonumber\\
\label{eq:9}
\end{eqnarray}
and the integration contour C$_+$ is defined subsequently.
The Green's function ${\cal G}_\kappa^{ij}(x_2,x_1,z)$ in
Eq.~(\ref{eq:2}) has poles along the real axis in the range of
integration over $z$, and in the perturbation expansion of ${\cal
G}_\kappa^{ij}(x_2,x_1,z)$, in powers of a perturbing potential $\delta
V$ and the unperturbed Green's function $G_\kappa^{ij}(x_2,x_1,z)$,
higher order poles are introduced as noted in Sec. \ref{sec:pert}.
Here, our method of isolating those poles and evaluating their
contribution is described. 
In terms of the spectral resolution of the unperturbed
radial Green's function
\begin{eqnarray}
G_\kappa^{ij}(x_2,x_1,z) = \sum_m\frac{f_{m,i}(x_2)f_{m,j}(x_1)
                            }{ E_m - z} \ ,
\label{eq:7}
\end{eqnarray}
Eq.~(\ref{eq:fgreencors}) reads
\begin{eqnarray}
&&\delta G_\kappa^{ij}(x_2,x_1,z) = - \int_0^\infty dx_3 \, x_3^2 \ 
\sum_{k=1}^2 \, \sum_{m_2,m_1}\nonumber \\
&&\quad\frac{f_{m_2,i}(x_2)f_{m_2,k}(x_3) }{ E_{m_2} - z}
\delta V(x_3) 
\frac{f_{m_1,k}(x_3)f_{m_1,j}(x_1) }{E_{m_1} - z} \ ,
\label{eq:8}
\end{eqnarray}
which explicitly shows the second and first-order poles.
In (\ref{eq:8}), only states $m_2, m_1$ with spin-angular momentum
quantum $\kappa$ contribute.
The principal parts of $\delta G_\kappa^{ij}(x_2,x_1,z)$ are identified 
by expanding the functions in Eq.~(\ref{eq:9}) in Laurent series about
each pole.  For $z\approx E_m$,
\begin{eqnarray}
G_\kappa^{ij}(x_2,x_1,z) &=& \frac{f_{m,i}(x_2)f_{m,j}(x_1)
                            }{ E_m - z}
   +G_\kappa^{R,ij}(x_2,x_1,E_m)\nonumber\\
&& + {\cal O}(z-E_m),
\label{eq:15}
\end{eqnarray}
where $G_\kappa^{R,ij}(x_2,x_1,E_m)$ are the radial components of the 
reduced Green's function given in Eq.~(\ref{eq:redgreen})
\begin{eqnarray}
G_\kappa^{R,ij}(x_2,x_1,E_m) &=& \lim_{z \rightarrow E_m}
\bigg[G_\kappa^{ij}(x_2,x_1,z) 
\nonumber\\
&&\qquad \qquad -\frac{f_{m,i}(x_2)f_{m,j}(x_1)
                            }{E_m - z}\bigg]
\nonumber\\
&=&\sum_{\stackrel{\scriptstyle l}{\scriptstyle E_l \ne E_m}}
\frac{f_{l,i}(x_2)f_{l,j}(x_1)}{ E_l - E_m} \ .
\label{eq:redgrcp}
\end{eqnarray}
Hence
\widetext
\begin{eqnarray}
\delta G_\kappa^{ij}(x_2,x_1,z) &=& - \int_0^\infty dx_3 \, x_3^2
\sum_{k=1}^2 \bigg[f_{m,i}(x_2)f_{m,k}(x_3)  \delta V(x_3) 
f_{m,k}(x_3)f_{m,j}(x_1) \frac{1}{(E_m - z)^2} 
\nonumber\\
&&\qquad\qquad\qquad\qquad+G_\kappa^{R,ik}(x_2,x_3,E_m) \delta V(x_3) f_{m,k}(x_3)f_{m,j}(x_1)
\frac{1}{E_m - z}
\nonumber\\
&&\qquad\qquad\qquad\qquad+ f_{m,i}(x_2)f_{m,k}(x_3) \delta V(x_3) G_\kappa^{R,kj}(x_3,x_1,E_m)
\frac{1}{E_m - z}
\bigg] + {\cal O}(1) \ ,
\label{eq:redgreenb}
\end{eqnarray}
or, in view of Eqs.~(\ref{eq:foen}) and (\ref{eq:radwfcorr}),
\begin{eqnarray}
\delta G_\kappa^{ij}(x_2,x_1,z) &=& - f_{m,i}(x_2)f_{m,j}(x_1) \frac{\delta E_m
                            }{ (E_m - z)^2}
+ \Big[ \delta f_{m,i}(x_2) \, f_{m,j}(x_1)
  + f_{m,i}(x_2) \, \delta f_{m,j}(x_1) \Big]\frac{1}{E_m - z}
+ {\cal O}(1) \ .
\end{eqnarray}
\narrowtext

The same result may be obtained by expanding the pole contribution to the 
full Green's function
\begin{eqnarray}
{\cal G}_\kappa(x_2,x_1,z) = 
\frac{{\cal F}_{m,i}(x_2){\cal F}_{m,j}(x_1)
                            }{ {\cal E}_m - z}
+{\cal O}(1)
\end{eqnarray}
in powers of $\delta V(x)$, with
\begin{eqnarray}
{\cal E}_m &=& E_m + \delta E_m +\cdots
\nonumber\\
{\cal F}_{m,i}(x) &=& f_{m,i}(x) + \delta f_{m,i}(x) +\cdots \ ,
\end{eqnarray}
and retaining only the first-order correction
\begin{eqnarray*}
\lefteqn{- f_{m,i}(x_2)f_{m,j}(x_1)\frac{\delta E_m
                            }{ (E_m - z)^2}} && \\
&+& \Big[ \delta f_{m,i}(x_2) \, f_{m,j}(x_1)
 + f_{m,i}(x_2) \, \delta f_{m,j}(x_1) \Big]\frac{1}{E_m - z} \ .
\end{eqnarray*}
In addition to the expansion of the Green's function correction,
we have
\begin{eqnarray}
A_\kappa^{ij}(x_2,x_1,z) &=& A_\kappa^{ij}(x_2,x_1,E_m)  \nonumber \\
 &&+(z-E_m)\,B_\kappa^{ij}(x_2,x_1,E_m)
 \nonumber \\
&& + {\cal O}\left((z-E_{m})^2\right) \ ,
\label{eq:17}
\end{eqnarray}
where
\begin{eqnarray}
B_\kappa^{ij}(x_2,x_1,E_m) = \frac{d}{ dz} \, \Big[A_\kappa^{ij}(x_2,x_1,z)\Big]_{z=E_m} \ .
\label{eq:18}
\end{eqnarray}
\widetext
The complete expansion for $z\approx E_m$ is
\begin{eqnarray}
&& \int_0^\infty dx_3 \, x_3^2 \, \sum_{k=1}^2
  G_\kappa^{ik}(x_2,x_3,z) \delta V(x_3) G_\kappa^{kj}(x_3,x_1,z) \, A_\kappa^{ij}(x_2,x_1,z)
\nonumber\\
&& \quad\qquad= f_{m,i}(x_2) \, f_{m,j}(x_1) \, \delta E_m \bigg[
A_\kappa^{ij}(x_2,x_1,E_m) \, \frac{1}{(E_m - z)^2}
-B_\kappa^{ij}(x_2,x_1,E_m)\frac{1}{E_m - z} \bigg]
\nonumber\\
&& \qquad\qquad\qquad\qquad\qquad\qquad-\Big[ \delta f_{m,i}(x_2) \, f_{m,j}(x_1)
 + f_{m,i}(x_2) \, \delta f_{m,j}(x_1) \Big]A_\kappa^{ij}(x_2,x_1,E_m)
 \frac{1}{E_m - z}
+{\cal O}(1) \ .
\label{eq:20}
\end{eqnarray}
\narrowtext

Our strategy for dealing with poles in the low-energy part is to 
calculate the line integral over $z$ of the difference between the 
complete integrand and the pole terms, and add the pole terms 
integrated analytically.  In the unperturbed self-energy calculation, 
the singularities along the real axis in the interval $(0,1)$ are 
poles, and the appropriate prescription for integration over $z$ 
yields the principal value integral in Eq.~(\ref{eq:lowe_ang}).  In the 
present context, there are double poles as well, so it is necessary to 
reexamine the original derivation to obtain the correct prescription.  
It follows from the discussion in Ref.~\cite{moh74a} that the 
integration over $z$ can be written as
\begin{eqnarray}
{\rm Re} \int_{\rm C_{+}}dz\, {\cal U}(z),
\label{eq:11}
\end{eqnarray} 
where C$_+$ is a contour that extends from $z=0$ to $z=E_n$ above the 
real axis in the complex $z$ plane. Here we use a method based on an 
analytic evaluation of the pole terms, as described in Ref.~\cite{mak92}. 
With the notation 
${\cal U}_{\rm p}$ for the pole terms of ${\cal U}$, we have
\begin{eqnarray}
{\rm Re} \int_{{\rm C}_+}dz\, {\cal U}(z) &= &{\rm Re} 
\int_{{\rm C}_+}dz\,\left[{\cal U}(z)-{\cal U}_{\rm p}(z)
+{\cal U}_{\rm p}(z)\right], \nonumber \\
&=&\int_0^{E_n}dz\Bigg[{\cal U}(z)-{\cal U}_{p}(z)+\nonumber \\ && 
\qquad\qquad\frac{1}{E_{n}}Re\int_{{\rm C}_+}dz^\prime\, 
{\cal U}_{\rm p}(z^\prime)\Bigg] \ .
\label{eq:polerem}
\end{eqnarray} 
The pole at $z = E_n$, the endpoint of the integral over $z$, does not cause 
any problem, as follows from the discussion in Sec.~\ref{sec:loenper}
[see Eqs.~(\ref{eq:eder}) to (\ref{eq:hamilcom})].
The relevant integrals for the analytic evaluation of the pole terms are
\begin{eqnarray}
{\rm Re} \int_{{\rm C}_+}dz\, \frac{1}{(E_m - z)^2} = - \frac{E_n}{E_m(E_n - E_m)} \ ,
\label{eq:12}
\end{eqnarray} 
\begin{eqnarray}
{\rm Re} \int_{{\rm C}_+}dz\, \frac{1}{E_m - z} 
= \ln \left[ \frac{E_m}{ E_n-E_m}\right] \ ,
\label{eq:13}
\end{eqnarray} 
where 
\begin{eqnarray}
0<E_m<E_n \ .
\end{eqnarray}

Applying these results to Eq.~(\ref{eq:lowedgang}), we have
\widetext
\begin{eqnarray}
\delta_{\rm G}  {\cal E}_{\rm L} &=& \frac{\alpha}{\pi} \, {\rm Re} 
\int_{{\rm C}_+}dz\,\delta_{\rm G} \, {\cal U}(z)  
\nonumber\\
 &=& \frac{\alpha}{\pi} \,\int_0^{E_n} 
\left\{\delta_{\rm G} \,{\cal U}(z) - 
\sum_{\stackrel{\scriptstyle m }{ \scriptstyle E_m < E_n}}
\left[\frac{R^{(2)}_{n,m}}{ (E_m-z)^2}+
\frac{R^{(1)}_{n,m}}{ (E_m-z)}\right]\right\}
\nonumber\\ && \qquad \qquad
+ \frac{\alpha}{\pi} \,
\sum_{\stackrel{\scriptstyle m}{\scriptstyle E_m < E_n}}
\left[- \frac{R^{(2)}_{n,m} \, E_n }{ E_m(E_n-E_m)}+
R^{(1)}_{n,m}\ln\left(\frac{E_m }{ E_n-E_m}\right)\right]
\end{eqnarray}
where
\begin{eqnarray}
R^{(2)}_{n,m} &=& \delta E_m
\int_0^\infty dx_2 \, x_2^2 \int_0^\infty dx_1 \, x_1^2
\sum_{i,j=1}^2
f_{n,\bar \imath}(x_2) f_{m,i}(x_2) \, f_{m,j}(x_1) \,
f_{n,\bar \jmath}(x_1) A_\kappa^{ij}(x_2,x_1,E_m)
\end{eqnarray}
and
\begin{eqnarray}
R^{(1)}_{n,m} &=& -\delta E_m
\int_0^\infty dx_2 \, x_2^2 \int_0^\infty dx_1 \, x_1^2
\sum_{i,j=1}^2
f_{n,\bar \imath}(x_2) f_{m,i}(x_2) \, f_{m,j}(x_1) \,
f_{n,\bar \jmath}(x_1) B_\kappa^{ij}(x_2,x_1,E_m)
\nonumber\\
&&-\int_0^\infty dx_2 \, x_2^2 \int_0^\infty dx_1 \, x_1^2
\sum_{i,j=1}^2
f_{n,\bar \imath}(x_2) \Big[ \delta f_{m,i}(x_2) \, f_{m,j}(x_1)
 + f_{m,i}(x_2) \, \delta f_{m,j}(x_1) \Big]
f_{n,\bar \jmath}(x_1) A_\kappa^{ij}(x_2,x_1,E_m)
\end{eqnarray}
\narrowtext

\subsection{Numerical evaluation of the first-order correction to the
 Green's function for the low-energy part}
\label{subsec:numdglow}
Numerical evaluation of the Coulomb Green's functions in this paper
is based on
the explicit formulas given, for example, in Eqs.~(A.16) and (A.17)
of Ref.~\cite{moh74a}, together with the numerical algorithms 
described in Ref.~\cite{moh74b}.
The first-order correction to the Green's function in 
Eq.~(\ref{eq:fgreencors}), for the range of arguments relevant to the 
low-energy part, is evaluated by numerical integration over $x_3$, 
where the interval of integration is divided into four subintervals:
$(0,y_1)$, $(y_1,y_2)$, $(y_2,y_3)$, and $(y_3,\infty)$. 
Defining $p= 2 \sqrt{1-z^2}$, $y_0=3$,
and assuming $x_2 \le x_1$, we choose $y_1=\min(y_0, p x_2)$,
$y_2=\min(\max(y_0, p x_2),p x_1)$, and $y_3=\max(y_0, p x_1)$.  
In the first interval we make the substitution $x_3=\xi^3$, and
integrate over $\xi$ by Gauss-Legendre quadrature with 12 to 26
integration points. In the second and third interval we also 
do Gauss-Legendre integration,
with 17 to 33 and 11 to 21 points, respectively. We use 6 to 18
point Gauss-Laguerre integration for the remaining interval. The
integrations over the 
second and fourth intervals are the least accurate. For a $2s$ electron
at $Z=20$ the integral over $(y_1,y_2)$ has an error of a few
parts in $10^{6}$ in the worst case.


\subsection{Reduced Green's Function}
\label{subsec:redgreen}
In the preliminary version of this calculation \cite{iam92}, a purely
numerical method of evaluating the reduced Green's function was
employed.  However, while that method is adequate at high $Z$, it
gives unsatisfactory results for the 2s state when $Z \le 20$, so a
new method was developed that yields better precision.  As a check of
the coding of the later method, the results of the two methods were
compared and are in agreement within a relative difference of
$10^{-6}$ over a wide range of the variables.  Both methods are
briefly described in the following subsections.

\subsubsection{Evaluation of the Reduced Green's Function by numerical pole removal}
Eq.~(\ref{eq:15}) can be written as
\begin{eqnarray}
G^{ij}_\kappa(x_2,x_1,E_n+\epsilon) &=& - \frac{f_{n,i}(x_2)f_{n,j}(x_1)}
{\epsilon} 
\nonumber \\
&& + G_\kappa^{{\rm R},ij}(x_2,x_1,E_n) + {\cal O}(\epsilon) \ .
\label{eq:redgreq}
\end{eqnarray}
As an immediate consequence of this relation, we have
\begin{eqnarray}
{1\over2}\big[
G_\kappa^{ij}(x_2,x_1,E_n+\epsilon)  +
G_\kappa^{ij}(x_2,x_1,E_n-\epsilon) 
\big]\qquad\nonumber\\
 = G_\kappa^{{\rm R},ij}(x_2,x_1,E_n) + {\cal O}(\epsilon^2) \ ,
\end{eqnarray}
so the reduced Green's function can be easily obtained 
from the full Green's function
by symmetric interpolation of the energy variable.
We form a linear combination of two such
interpolations in order to obtain a result with an error of order
$\epsilon^4$ rather than $\epsilon^2$.  In particular, we have
\begin{eqnarray}
&&\frac{q}{2}\big[G_\kappa^{ij}(x_2,x_1,E_n+\epsilon)
 + G_\kappa^{ij}(x_2,x_1,E_n-\epsilon)\big]
\nonumber\\
&&+\frac{(1-q)}{2}\big[G_\kappa^{ij}(x_2,x_1,E_n+\epsilon')
 + G_\kappa^{ij}(x_2,x_1,E_n-\epsilon')\big] 
\nonumber\\
&& = G_\kappa^{{\rm R},ij}(x_2,x_1,E_n)
+ C_2 \left[q \, \epsilon^2 + (1-q) \,{\epsilon'}^2\right]
\nonumber\\
&&\quad + C_4 \left[q \, \epsilon^4 + (1-q) \,{\epsilon'}^4\right]
 + {\cal O}(\epsilon^6) + {\cal O}({\epsilon'}^6) \ ,
\label{eq:23}
\end{eqnarray}
where the choice
\begin{eqnarray}
\epsilon' = \left(\frac{q}{q-1}\right)^\frac{1}{2}\epsilon; \qquad\qquad q > 1
\label{eq:24}
\end{eqnarray}
provides
\begin{eqnarray}
q \, \epsilon^2 + (1-q) \,{\epsilon'}^2 = 0
\label{eq:25}
\end{eqnarray}
with the parameter $q$ free to vary in the range $(1,\infty)$.
The choice $q = \frac{4}{3}$ gives the Lagrange interpolation formula
with equally spaced evaluation points:
$\epsilon' = 2 \epsilon$.  
With the choice in Eq.~(\ref{eq:24}), the
fourth-order term is proportional to
\begin{eqnarray}
\left| q \, \epsilon^4 + (1-q) \,{\epsilon'}^4 \right|
= \frac{q}{q-1} \, \epsilon^4 \ .
\end{eqnarray}
The coefficient of $\epsilon^4$ is $4$ for equally spaced points and 
approaches $1$ as $q \rightarrow \infty$.  
As a compromise between a large coefficient for 
$q$ near 1 and the minimum coefficient as $q \rightarrow \infty$, with 
a correspondingly larger roundoff error, we employ the value $q = 10$.  
The interpolation interval is taken to be $\epsilon
= 0.02\left(E_{n+1}-E_{n}\right)$ 
where $E_n$ and $E_{n+1}$ are the Dirac eigenvalues
for principal quantum number $n$ and $n+1$ for the same $\kappa$.
This interval avoids overlap with the nearest 
pole of the Green's function.

\subsubsection{Direct evaluation of the Reduced Green's Function}
The method developed for the present work
is similar to that of Hylton \cite{hyl84,hyl85},
but differs in the details of its implementation.
From Eq.~(\ref{eq:redgreq}), it is evident that we obtain the reduced
Green's function by expanding the various components in the 
explicit expression for the
radial Green's function in powers of $\epsilon = z - E_n$ and 
keeping only the final combinations of terms that are of order $1$.
To implement this, at each step in the numerical
evaluation of the reduced Green's function we
calculate only the coefficients of the leading two terms in the power
series in $\epsilon$ and discard the higher-order terms.
In certain cases, it is necessary to begin with
three terms in the expansions,
because the leading term either vanishes or cancels an equal 
leading term in forming a difference.  
As suggested by the form of the following equations, many of
the coefficients follow
from combinations of coefficients that appear earlier in the
calculation.

The code for the numerical calculation was written by
modifying the existing code for the radial Green's
function described in 
Refs. \cite{moh74b} and \cite{mps98},
so only a few details that illustrate the approach
are given here.
We define the radial quantum number $n_{\rm r} = n - |\kappa|$.
Expansions are needed for $z$:
\begin{eqnarray}
z &=& z_0 + z_1\epsilon + z_2\epsilon^2 +\cdots \nonumber\\
&&z_0 = E_n; \
z_1 = 1; \
z_2 = 0 \ ,
\end{eqnarray}
for $c=\sqrt{1-z^2}$:
\begin{eqnarray}
c &=& c_0 + c_1\epsilon + c_2\epsilon^2 +\cdots \nonumber\\
&&c_0 =\frac{\gamma E_n}{n_r + \lambda}; \
c_1 = - \frac{n_{\rm r}+\lambda}{\gamma}; \
c_2 = -\frac{1}{2c_0^3} \ ,
\end{eqnarray}
where $\gamma = Z\alpha$ and $\lambda = \sqrt{\kappa^2-\gamma^2}$,
for $w = \gamma/c$:
\begin{eqnarray}
w &=& w_0 + w_1\epsilon + w_2\epsilon^2 +\cdots \nonumber\\
&&w_0 = \frac{\gamma}{c_0}; \
w_1 = -\frac{\gamma c_1}{c_0^2}; \
w_2 = \frac{\gamma(c_1^2-c_0c_2)}{c_0^3} \ ,
\end{eqnarray}
and for $\nu = \gamma z/c$:
\begin{eqnarray}
\nu &=& \nu_0 + \nu_1\epsilon + \nu_2\epsilon^2 +\cdots \nonumber\\
&&\nu_0 = z_0 w_0; \
\nu_1 = z_0 w_1 + z_1 w_0; \nonumber\\
&&\nu_2 = z_0 w_2 + z_1 w_1 \ .
\end{eqnarray}
Expressions that appear in the definitions of the
radial Green's functions include $\lambda - \nu$:
\begin{eqnarray}
\lambda - \nu &=& -n_{\rm r} 
- \nu_1\epsilon -\nu_2\epsilon^2 + \cdots \nonumber \\
&=&\Big\{\begin{array}{ll}
- \nu_1\epsilon -\nu_2\epsilon^2 + \cdots
& {\rm if} \ n_{\rm r} \ = 0 \\
-n_{\rm r} - \nu_1\epsilon + \cdots 
& {\rm if} \ n_{\rm r} \ \ne 0
         \end{array}
\end{eqnarray}
and ${\it \Gamma}(\lambda - \nu)$:
\begin{eqnarray}
{\it \Gamma}(\lambda - \nu) &=&
{\it \Gamma}^{(n_{\rm r})}_{-1} \epsilon^{-1} 
+ {\it \Gamma}^{(n_{\rm r})}_0  + \cdots
\nonumber\\
&&{\it \Gamma}^{(0)}_{-1} = - {1 \over \nu_1}; \ 
{\it \Gamma}^{(0)}_{0} =  {\nu_2 \over \nu_1^2} - \gamma_{\rm E}
\nonumber\\
&&{\it \Gamma}^{(n_{\rm r})}_{-1} 
= - {1\over n_{\rm r}}{\it \Gamma}^{(n_{\rm r}-1)}_{-1}
\nonumber\\
&&{\it \Gamma}^{(n_{\rm r})}_{0} 
= - {1\over n_{\rm r}}{\it \Gamma}^{(n_{\rm r}-1)}_{0}
+ {\nu_1\over n_{\rm r}^2}{\it \Gamma}^{(n_{\rm r}-1)}_{-1} \ ,
\end{eqnarray}
where $\gamma_{\rm E} = 0.577\dots$ is Euler's constant.
The recursion relations used to calculate the power series 
for the Whittaker functions are treated in a similar manner.
For example, for the power series evaluation of 
$M_{\nu+(1/2),\lambda}(2cx)$ by means of the recursion relations
in Eq.~(D.2) of \cite{moh74b}, we write
\begin{eqnarray}
T(m) &=& T_0(m) + T_1(m)\epsilon + \cdots 
\nonumber\\
\vbox to 15pt {}
&&T_0(0) = 1; \ T_1(0) = 0
\nonumber\\
\vbox to 20pt {}
T_0(m+1) &=& 
{(m-n_{\rm r})2c_0x \over (m + 2\lambda + 1)(m + 1)}T_0(m)
\nonumber\\
\vbox to 20pt {}
T_1(m+1) &=& 
{(m-n_{\rm r})2c_0x \over (m + 2\lambda + 1)(m + 1)}T_1(m)
\nonumber\\
&&+ {[(m-n_{\rm r})c_1-\nu_1c_0]2x 
\over (m + 2\lambda + 1)(m + 1)}T_0(m)
\label{eq:mfun}
\end{eqnarray}
The termination of the power series for the
leading term in Eq.~(\ref{eq:mfun})
at $m = n_{\rm r}$ corresponds to fact 
that the leading term in Eq.~(\ref{eq:redgreq}) is 
proportional to the bound-state wave function.

The calculation of the reduced Green's function in this 
work is based on the application of two-term expansions in 
$\epsilon$, as described above, to the
numerical evaluation of the complete Green's function 
(see Appendix D of Ref.~\cite{moh74b}).  The numerical 
value of the reduced Green's function is just the 
collection of terms with combined order 1 in $\epsilon$.

\subsection{Numerical evaluation of the first-order correction to the 
wave function}
\label{sec:numedphi}
The first-order correction to the wave function, given by 
Eq.~(\ref{eq:radwfcorr}), is evaluated with the 
aid of the reduced Green's function as 
described in Sec.~\ref{subsec:redgreen}.
The numerical integration in Eq.~(\ref{eq:radwfcorr}) is divided
into 3 segments:
\widetext
\begin{eqnarray}
\delta f_{n,i}(x)&=&\frac{1}{p_n}\Bigg[5\int_0^{1} 
du \,  x_1 u^4 \left( x_1u^5/p_n\right)^2 
\sum_{j=1}^{2}G_{\kappa}^{R,ij}(x, x_1u^5/p_n,E_n)
\delta V(x_1u^5/p_n) f_{n,j}({ x_1u^5/p_n})
\nonumber\\
&&+\int_{x_1}^{x_2} d y  \left({y/p_n}\right)^2 
\sum_{j=1}^{2}
G_{\kappa}^{R,ij}(x,{y/p_n},E_n) \delta V({y/p_n}) f_{n,j}({y/p_n})
\nonumber\\
&&+\int_{x_2}^{\infty} d y \left({y/p_n}\right)^2 \sum_{j=1}^{2}
G_{\kappa}^{R,ij}(x,{y/p_n},E_n) \delta V({y/p_n}) f_{n,j}({y/p_n})
\Bigg] \ ,
\label{eq:pertwf}
\end{eqnarray}
\narrowtext\noindent
where $p_n=2\sqrt{1-E_n^2}$ is the coefficient of the argument in
the exponent that governs the
behavior of the integrand for large values of the argument, 
$x_1=\min\left(xp,3\right)$, and
$x_2=\max\left(xp,3\right)$. The first and second integrals are evaluated by
means of 30 point Gauss-Legendre quadrature and the third is evaluated with
15 point Gauss-Laguerre quadrature.  The accuracy of the perturbed wave function
calculated according to Eq.~(\ref{eq:pertwf})
has been tested in the Coulomb case by comparison to the result
obtained by numerical differentiation of the Coulomb wave function.

For values of the argument of the first-order correction to 
the wave function near the origin, 
we found that greater numerical accuracy and speed could be obtained
with a numerical evaluation based on the expansion in powers of $x$.
This expansion is described in Appendix \ref{app:orexpwf1}.


\section{High energy part}
\label{sec:high_e}
The high-energy part, given by the integral
in Eq.~(\ref{eq:high_e}), must be regularized, since it is formally
infinite. We employ the Pauli-Villars regularization scheme, 
following the method of Refs.~\cite{iam92,iam98} to isolate and
remove the divergent contributions. In Ref.~\cite{iam98} we 
demonstrated that suitable numerical convergence can be achieved through
the use of a term-by-term subtraction method.  This particular method 
has the advantage that it does not require a mix of coordinate-space
and momentum-space calculations as do earlier methods, but works
entirely within coordinate space.  In this method, the high-energy 
part $ {\cal E}_{\rm H}$, given by Eq.~(\ref{eq:high_e}), 
is separated into two
parts: ${\cal E}_{\rm HA}$ and ${\cal E}_{\rm HB}$. The divergences
are all contained in ${\cal E}_{\rm HA}$ and can be calculated
completely analytically, while ${\cal E}_{\rm HB}$ is
finite and is treated numerically. In this section we describe the 
method used to evaluate $\delta E_{\rm HB}$,
while the method to compute $\delta E_{\rm HA}$ is discussed in
Sec.~\ref{sec:an_sub}; the total is 
\begin{eqnarray}
\delta E_{\rm H} = \delta E_{\rm HA} + \delta E_{\rm HB} \ .
\end{eqnarray}.

The high-energy remainder with term-by-term subtraction, from Eq.~(32)
in \cite{iam98}, is written as
\widetext
\begin{eqnarray}
{\cal E}_{\rm HB}&=& {\cal  E}_{\rm H} - {\cal E}_{\rm HA}
\nonumber \\
&=&\frac{\alpha}{ 2 \pi {\rm i}} \int_{\rm C_H}
dz\ \int_0^\infty dx_2 \, x_2^2 \int_0^\infty dx_1 \, x_1^2 
\nonumber \\
&&\times\bigg\{\sum_{|\kappa|= 1}^{\infty}
\bigg[{\cal K}_\kappa(x_2,x_1,z) -
{\cal K}^{(0,0)}_\kappa(x_2,x_1,z) -
{\cal K}^{(0,1)}_\kappa(x_2,x_1,z) -
{\cal K}^{(1,0)}_\kappa(x_2,x_1,z)\bigg]
- {\cal K}^{(0,2)}_{\rm D}(x_2,x_1,z) \bigg\} \ ,
\label{eq:hb}
\end{eqnarray}
with 
\begin{eqnarray}
{\cal K}_\kappa(x_2,x_1,z) 
 =
\sum_{i,j=1}^2\big[{\cal F}_{n,i}(x_2)
{\cal G}_\kappa^{ij}(x_2,x_1,z){\cal F}_{n,j}(x_1) A_\kappa(x_2,x_1)
-{\cal F}_{n,\bar \imath}(x_2){\cal G}_\kappa^{ij}(x_2,x_1,z)
  {\cal F}_{n,\bar \jmath}(x_1)A^{ij}_\kappa(x_2,x_1)\big] \ ,
\label{eq:Kk}
\end{eqnarray}
\begin{eqnarray}
{\cal K}^{(0,0)}_\kappa(x_2,x_1,z)=  A_\kappa
\left\{F^{11}_\kappa(x_2,x_1,z)
\left[{\cal F}_{n,1}^2(x_2)-3{\cal F}_{n,2}^2(x_2)\right]
+F^{22}_{-\kappa}(x_2,x_1,z)
\left[{\cal F}_{n,2}^2(x_2)-3{\cal F}_{n,1}^2(x_2)\right]\right\} \ ,
\label{eq:K00k}
\end{eqnarray}
\begin{eqnarray}
&&{\cal K}^{(0,1)}_\kappa(x_2,x_1,z)
\nonumber\\
&&\quad= \left\{
F^{11}_\kappa(x_2,x_1,z)
\left[{\cal F}_{n,1}(x_2)\left(
B_\kappa\frac{\partial}{\partial x_2}
+C_\kappa\frac{1+\kappa_n}{ x_2}\right)
{\cal F}_{n,1}(x_2)-{\cal F}_{n,2}(x_2)
\left(3B_\kappa\frac{\partial}{\partial x_2}
-C_\kappa\frac{1-\kappa_n}{ x_2}\right){\cal F}_{n,2}(x_2)
\right]\right.
\nonumber\\
&&\qquad+F^{22}_{-\kappa}(x_2,x_1,z)
\left[{\cal F}_{n,2}(x_2)
\left( B_\kappa\frac{\partial}{\partial x_2}
+C_\kappa\frac{1-\kappa_n}{ x_2}\right)
{\cal F}_{n,2}(x_2)
-{\cal F}_{n,1}(x_2)
\left(3B_\kappa\frac{\partial}{\partial x_2}
-C_\kappa\frac{1+\kappa_n}{ x_2}\right)
{\cal F}_{n,1}(x_2)\right]
\nonumber\\
&&\qquad+
2 \, F^{12}_{-\kappa}(x_2,x_1,z)
{\cal F}_{n,1}(x_2) \left(
B_\kappa\frac{\partial}{\partial x_2}
+C_\kappa\frac{1-\kappa_n}{ x_2}\right)
{\cal F}_{n,2}(x_2)
\nonumber\\&&\qquad\left. +
2 \, F^{21}_{\kappa}(x_2,x_1,z)
{\cal F}_{n,2}(x_2) \left(
B_\kappa\frac{\partial}{\partial x_2}
+C_\kappa\frac{1+\kappa_n}{ x_2}\right)
{\cal F}_{n,1}(x_2)\right\} \ ,
\label{eq:K01k}
\end{eqnarray}
and 
\begin{eqnarray}
{\cal K}^{(1,0)}_\kappa(x_2,x_1,z)&= -
{\cal V}(x_2)A_\kappa&\left\{\frac{d}{ dz}F^{11}_\kappa(x_2,x_1,z)
\left[{\cal F}_{n,1}^2(x_2)-3{\cal F}_{n,2}^2(x_2)\right]\right.
\nonumber\\&&\, \left.
+\frac{d}{ dz}F^{22}_{-\kappa}(x_2,x_1,z)
\left[{\cal F}_{n,2}^2(x_2)-3{\cal F}_{n,1}^2(x_2)\right]\right\} \ .
\label{eq:K10k}
\end{eqnarray}
\narrowtext
In Eqs.~(\ref{eq:Kk}) to (\ref{eq:K10k}) $A^{ij}_\kappa$, $A_\kappa$,
$B_\kappa$ and $C_\kappa$ are integrals over coordinate directions 
defined in Refs.\
\cite{moh74a,iam98,mak92,moh82}, ${\cal F}_{n,i}(x)$ are radial 
components of the bound-state wave function as before, 
and $\kappa_n$ is the spin-angular momentum quantum number of the 
bound state $n$.  The expressions for the free Green's function
radial components $F^{ij}_{\kappa}(x_2,x_1,z)$ and their derivatives
can be found in \cite{iam98} and those of the Coulomb Green's function
can be found in \cite{moh74a}. The methods we used for summation over
angular momentum $\kappa$ and for numerical integrations are identical
to those described in \cite{iam98} and will not be repeated here. We
also found that the convergence of the numerical integration was much
better than in the case of the unperturbed self-energy. We thus did
not use the extra subtraction term $K^{(0,2)}_{\rm D}(x_2,x_1,z)$ that
was necessary in \cite{iam98} to obtain good convergence at low
$Z$. 
The high-energy remainder for the self-energy screening is obtained
from Eq.~(\ref{eq:hb}) as described in Sec.~\ref{sec:pert}
as the sum $\delta_{\phi} {\cal E}_{\rm HB}+\delta_{\rm
G} {\cal E}_{\rm HB}+\delta_{\rm E} {\cal E}_{\rm HB}$. In the three
following subsections we derive the expressions that are used to
obtain $\delta_{\phi} {\cal E}_{\rm HB}$, $\delta_{\rm G} {\cal
E}_{\rm HB}$, and $\delta_{\rm E} {\cal E}_{\rm HB}$ from 
Eqs.~(\ref{eq:hb}) to (\ref{eq:K10k}).

\subsection{Wave function correction}
\label{sec:wf_high}
To obtain the expression for the high-energy remainder for the wave
function correction, we need the functional derivatives of
Eqs.~(\ref{eq:Kk}) to (\ref{eq:K10k}) with respect to
the radial wave functions ${\cal F}_{n,i}(x)$.  For the full expression
(\ref{eq:Kk}) we have
\widetext
\begin{eqnarray}
K_{\phi,\kappa}(x_2,x_1,z) 
 &=&
\sum_{i,j=1}^2\Big\{\left[f_{n,i}(x_2)G_\kappa^{ij}(x_2,x_1,z)\,\delta f_{n,j}(x_1)
+\delta f_{n,i}(x_2)G_\kappa^{ij}(x_2,x_1,z) f_{n,j}(x_1)\right]A_\kappa(x_2,x_1) 
\nonumber \\ && \qquad
-\left[ f_{n,{\bar \imath}}(x_2)G_\kappa^{ij}(x_2,x_1,z) \,
  \delta f_{n,{\bar \jmath}}(x_1)
+\delta f_{n,{\bar \imath}}(x_2)G_\kappa^{ij}(x_2,x_1,z)
   f_{n,{\bar \jmath}}(x_1)\right] A^{ij}_\kappa(x_2,x_1)\Big\} \ ,
\label{eq:Kkdf}
\end{eqnarray}
and for the subtraction terms, we obtain
\begin{eqnarray}
K^{(0,0)}_{\phi,\kappa}(x_2,x_1,z)&= &2 A_\kappa
\Big\{F^{11}_\kappa(x_2,x_1,z)
\left[f_{n,1}(x_2)\,\delta f_{n,1}(x_2)-3f_{n,2}(x_2)\,\delta f_{n,2}(x_2)\right]
\nonumber \\ && \qquad 
+F^{22}_{-\kappa}(x_2,x_1,z)
\left[f_{n,2}(x_2)\,\delta f_{n,2}(x_2)-3f_{n,1}(x_2)\,\delta f_{n,1}(x_2)\right]\Big\} \ ,
\label{eq:K00kdf}
\end{eqnarray}
\begin{eqnarray}
&&K^{(0,1)}_{ \phi,\kappa}(x_2,x_1,z)
\nonumber\\
&&\quad= \bigg\{
F^{11}_\kappa(x_2,x_1,z)
\bigg[
f_{n,1}(x_2)\left(
B_\kappa\frac{\partial}{\partial x_2}
+C_\kappa\frac{1+\kappa_n}{ x_2}
\right)\delta f_{n,1}(x_2)
-f_{n,2}(x_2)\left(3B_\kappa\frac{\partial}{\partial x_2}
-C_\kappa\frac{1-\kappa_n}{ x_2}\right)\delta f_{n,2}(x_2)
\nonumber\\&&\qquad\qquad\qquad\qquad\quad+
\delta f_{n,1}(x_2)\left(
B_\kappa\frac{\partial}{\partial x_2}
+C_\kappa\frac{1+\kappa_n}{ x_2}\right)f_{n,1}(x_2)
-\delta f_{n,2}(x_2)\left(3B_\kappa\frac{\partial}{\partial x_2}
-C_\kappa\frac{1-\kappa_n}{ x_2}\right)f_{n,2}(x_2)
\bigg]
\nonumber\\
&&\qquad+F^{22}_{-\kappa}(x_2,x_1,z)
\bigg[
f_{n,2}(x_2)\left( B_\kappa\frac{\partial}{\partial x_2}
+C_\kappa\frac{1-\kappa_n}{ x_2}\right)\delta f_{n,2}(x_2)
-f_{n,1}(x_2)\left(3B_\kappa\frac{\partial}{\partial x_2}
-C_\kappa\frac{1+\kappa_n}{ x_2}\right)\delta f_{n,1}(x_2)
\nonumber\\&&\qquad\qquad\qquad\qquad\quad+
\delta f_{n,2}(x_2)\left( B_\kappa\frac{\partial}{\partial x_2}
+C_\kappa\frac{1-\kappa_n}{ x_2}\right)f_{n,2}(x_2)
-\delta f_{n,1}(x_2)\left(3B_\kappa\frac{\partial}{\partial x_2}
-C_\kappa\frac{1+\kappa_n}{ x_2}\right)f_{n,1}(x_2)
\bigg]
\nonumber\\
&&\qquad+
2 \, F^{12}_{-\kappa}(x_2,x_1,z)
\bigg[
f_{n,1}(x_2) \left(
B_\kappa\frac{\partial}{\partial x_2}
+C_\kappa\frac{1-\kappa_n}{ x_2}\right)
\delta f_{n,2}(x_2)
+\delta f_{n,1}(x_2) \left(
B_\kappa\frac{\partial}{\partial x_2}
+C_\kappa\frac{1-\kappa_n}{ x_2}\right)
f_{n,2}(x_2)
\bigg]
\nonumber\\&&\qquad +
2 \, F^{21}_{\kappa}(x_2,x_1,z)
\bigg[
f_{n,2}(x_2) \left(
B_\kappa\frac{\partial}{\partial x_2}
+C_\kappa\frac{1+\kappa_n}{ x_2}\right)
\delta f_{n,1}(x_2)
+\delta f_{n,2}(x_2) \left(
B_\kappa\frac{\partial}{\partial x_2}
+C_\kappa\frac{1+\kappa_n}{ x_2}\right)
f_{n,1}(x_2)
\bigg]
\bigg\} \ ,
\label{eq:K01kdf}
\end{eqnarray}
and 
\begin{eqnarray}
K^{(1,0)}_{ \phi,\kappa}(x_2,x_1,z)&= -2
V\!(x_2)A_\kappa&\left\{\frac{d}{ dz}F^{11}_\kappa(x_2,x_1,z)
\left[f_{n,1}(x_2)\,\delta f_{n,1}(x_2)-3f_{n,2}(x_2)\,\delta f_{n,2}(x_2)\right]\right.
\nonumber\\&&\, \left.
+\frac{d}{ dz}F^{22}_{-\kappa}(x_2,x_1,z)
\left[f_{n,2}(x_2)\,\delta f_{n,2}(x_2)-3f_{n,1}(x_2)\,\delta f_{n,1}(x_2)\right]\right\} \ .
\label{eq:K10kdf}
\end{eqnarray}
In terms of the expressions in Eqs.~(\ref{eq:Kkdf}) to (\ref{eq:K10kdf}),
the first-order wave function correction to ${\cal E}_{\rm HB}$ is
\begin{eqnarray}
\delta_{\phi} {\cal E}_{\rm HB}&=&\frac{\alpha}{ 2 \pi {\rm i}} \int_{\rm C_H}
dz\ \int_0^\infty dx_2 \ x_2^2 \int_0^\infty dx_1 \ x_1^2 
\nonumber \\
&&\times\sum_{|\kappa|= 1}^{\infty}
\bigg[K_{\phi,\kappa}(x_2,x_1,z) -
K^{(0,0)}_{\phi,\kappa}(x_2,x_1,z) -
K^{(0,1)}_{\phi,\kappa}(x_2,x_1,z) -
K^{(1,0)}_{\phi,\kappa}(x_2,x_1,z)\bigg] \ .
\label{eq:hbdf}
\end{eqnarray}
\narrowtext

In order to evaluate the expression in (\ref{eq:K01kdf}), we need the
derivative of the bound-state Dirac wave function and the derivative
of its first-order correction in the potential $\delta V$. 
The differential equations for the large and small components of the 
unperturbed wave function are (see, e.g., \cite{iam92}, Appendix~A)
\begin{eqnarray}
      \frac{d}{d x} f_{n,1}(x)&=&-\frac{1+\kappa_n}{x}f_{n,1}(x)+\left[1+E_n-V(x)\right]f_{n,2}(x)
\nonumber\\
      \frac{d}{d x} f_{n,2}(x)&=&\left[1-E_n+V(x)\right]f_{n,1}(x)-\frac{1-\kappa_n}{x}f_{n,2}(x) \ ,
\nonumber\\
\label{eq:difdirac}
\end{eqnarray}
which yield the wave function derivatives from the
analytic expressions for the wave function.
We obtain analogous expressions for the perturbation of the
wave-function components in the potential $V(x)+\delta V(x)$. 
Retaining only first-order terms in
$\delta V(x)$, we obtain
\widetext
\begin{eqnarray}
      \frac{d}{d x} \delta f_{n,1}(x)&=&-\frac{1+\kappa_n}{x}\delta f_{n,1}(x)
+[1+E_n-V(x)]\,\delta f_{n,2}(x)
+[\delta E_n-\delta V(x)]f_{n,2}(x)
\nonumber\\
      \frac{d}{d x} \delta f_{n,2}(x)&=&[1-E_n+V(x)]\,\delta f_{n,1}(x)
-\frac{1-\kappa_n}{x}\delta f_{n,2}(x)
+[\delta V(x)-\delta E_n]f_{n,1}(x) \ .
\label{eq:fodifdirac}
\end{eqnarray}
\narrowtext

\subsection{Green's function correction}
\label{sec:green_high}
We do the corresponding calculation for the high-energy term to
account for variation of the Coulomb Green's function under a change
of the potential.  Since the Coulomb Green's function has no poles on the
high energy integration contour (which lies on the imaginary axis), we may
directly apply Eq.~(\ref{eq:fgreencors}) to obtain
\widetext
\begin{eqnarray}
K_{{\rm G},\kappa}(x_2,x_1,z) 
 & = &
-\sum_{i,j,k=1}^2 \int_0^\infty dx_3 \, x_3^2 \, \delta V(x_3)
 \bigg[f_{n,i}(x_2)G_\kappa^{ik}(x_2,x_3,z) G_\kappa^{kj}(x_3,x_1,z)f_{n,j}(x_1)A_\kappa(x_2,x_1)
\nonumber \\ &&\qquad \qquad \qquad\qquad
-f_{n,{\bar \imath}}(x_2)G_\kappa^{ik}(x_2,x_3,z) G_\kappa^{kj}(x_3,x_1,z)
  f_{n,{\bar \jmath}}(x_1)A^{ij}_\kappa(x_2,x_1)\bigg] \ .
\label{eq:Kkdg}
\end{eqnarray}
Only $K^{(1,0)}$ contributes to the subtraction term, and
we thus obtain from Eq.~(\ref{eq:K10k}) 
\begin{eqnarray}
&&K^{(1,0)}_{{\rm G},\kappa}(x_2,x_1,z)
\nonumber\\
&&\quad= - \delta V(x_2)A_{\kappa}
\bigg\{\frac{d}{ dz}F^{11}_\kappa(x_2,x_1,z)
\left[f_{n,1}^2(x_2)-3f_{n,2}^2(x_2)\right]
+\frac{d}{ dz}F^{22}_{-\kappa}(x_2,x_1,z)
\left[f_{n,2}^2(x_2)-3f_{n,1}^2(x_2)\right]\bigg\} \ .
\label{eq:K10kdg}
\end{eqnarray}
The Green's function correction to ${\cal E}_{\rm HB}$ is thus
\begin{eqnarray}
\delta_{\rm G} {\cal E}_{\rm HB}=\frac{\alpha}{ 2 \pi {\rm i}} \int_{\rm C_H}
dz\ \int_0^\infty dx_2 \ x_2^2 \int_0^\infty dx_1 \ x_1^2 
\sum_{|\kappa|= 1}^{\infty}
\bigg[K_{{\rm G},\kappa}(x_2,x_1,z) -
K^{(1,0)}_{{\rm G},\kappa}(x_2,x_1,z)\bigg]
\label{eq:hbdg}
\end{eqnarray}
\narrowtext

\subsection{Numerical evaluation of the first-order correction to the 
Green's function for the high-energy part}
\label{subsec:numdghigh}
The evaluation of the the Green's function correction to the 
high-energy part is the most difficult and time consuming part of the 
present calculation because of the additional non-trivial integration 
over $x_3$ and the necessity of evaluating integrals over a product of
two Coulomb Green's functions in Eq.~(\ref{eq:Kkdg}).
The calculation is facilitated somewhat by subtracting and adding an
additional term that closely approximates the term in 
Eq.~(\ref{eq:Kkdg}).
We write 
\widetext
\begin{eqnarray}
\lefteqn{
\int_0^\infty dx_3 \, x_3^2 \,\delta V(x_3)\, G_\kappa^{ik}(x_2,x_3,z) \, 
G_\kappa^{kj}(x_3,x_1,z)} \qquad && \nonumber \\
&=& \int_0^\infty dx_3 \, x_3^2\left[\delta V(x_3)-\frac{1}{2}\delta 
V(x_1)-\frac{1}{2}\delta V(x_2)\right] G_\kappa^{ik}(x_2,x_3,z) \,
G_\kappa^{kj}(x_3,x_1,z) 
 \nonumber \\&& \qquad+\frac{1}{2}\left[\delta V(x_1)+\delta V(x_2)\right]
\int_0^\infty dx_3 \, x_3^2  \,
G_\kappa^{ik}(x_2,x_3,z) \, G_\kappa^{kj}(x_3,x_1,z) \ ,
\label{eq:greengvg}
\end{eqnarray}
where (see Eq.~(35) from Ref.~\cite{iam92})
\begin{eqnarray}
\sum_{k=1}^2
\int_0^\infty dx_3 \, x_3^2  \,
G_\kappa^{ik}(x_2,x_3,z) \, G_\kappa^{kj}(x_3,x_1,z)
=
\frac{\partial}{\partial z}G_\kappa^{ij}(x_2,x_1,z) \ .
\end{eqnarray}
We calculate separately the two terms in the sum $\delta_{\rm G} {\cal E}_{\rm HB}
= \delta_{\rm G} {\cal E}_{\rm HB}^{(1)} + \delta_{\rm G} {\cal E}_{\rm HB}^{(2)}$, where
\begin{eqnarray}
\delta_{\rm G} {\cal E}_{\rm HB}^{(1)}&=&\frac{\alpha}{ 2 \pi {\rm i}} \int_{\rm C_H}
dz \int_0^\infty dx_2 \, x_2^2 \int_0^\infty dx_1 \, x_1^2 
\sum_{|\kappa|= 1}^{\infty}
\bigg\{-\frac{1}{2}\sum_{i,j=1}^2 \left[\delta
V(x_1)+\delta V(x_2)\right] \nonumber
 \\ &&\times
\left[f_{n,i}(x_2)\frac{\partial}{\partial z}G_\kappa^{ij}(x_2,x_1,z) f_{n,j}(x_1)A_\kappa(x_2,x_1)
-f_{n,{\bar \imath}}(x_2)\frac{\partial}{\partial z}G_\kappa^{ij}(x_2,x_1,z)
  f_{n,{\bar \jmath}}(x_1)A^{ij}_\kappa(x_2,x_1)\right] 
\nonumber\\&&
- K^{(1,0)}_{{\rm G},\kappa}(x_2,x_1,z)\bigg\}
\label{eq:hbdgderi}
\end{eqnarray}
and
\begin{eqnarray}
\delta_{\rm G} {\cal E}_{\rm HB}^{(2)}&=&\frac{\alpha}{ 2 \pi {\rm i}} \int_{\rm C_H}
dz\,\int_0^\infty dx_2 \, x_2^2 \int_0^\infty dx_1 \, x_1^2 
\int_0^\infty dx_3 \, x_3^2 
\left[-\delta V(x_3)+\frac{1}{2}\delta V(x_1)+\frac{1}{2}\delta V(x_2)\right] 
 \nonumber \\ && \times
\sum_{|\kappa|= 1}^{\infty}\sum_{i,j,k=1}^2 
\big[f_{n,i}(x_2)\,G_\kappa^{ik}(x_2,x_3,z) \,G_\kappa^{kj}(x_3,x_1,z)f_{n,j}(x_1)A_\kappa(x_2,x_1)
\nonumber\\
&&\qquad\qquad\qquad -f_{n,{\bar \imath}}(x_2)\,G_\kappa^{ik}(x_2,x_3,z) \,G_\kappa^{kj}(x_3,x_1,z)
  f_{n,{\bar \jmath}}(x_1)A^{ij}_\kappa(x_2,x_1)\big] \ .
\label{eq:kdiff}
\end{eqnarray}
The derivative of the Green's function in Eq.~(\ref{eq:hbdgderi}) is evaluated numerically
with a two-point formula:
\begin{equation}
        \frac{\partial}{\partial z}G_\kappa^{ij}(x_2,x_1,z) \approx \frac{
        G_\kappa^{ij}(x_2,x_1,z(1+\epsilon))-G_\kappa^{ij}(x_2,x_1,z(1-\epsilon))}
        {2z\epsilon} \ .
        \label{eq:gderi}
\end{equation}
\narrowtext

\subsection{Energy correction}
\label{sec:e_high}
We evaluate $\delta_{\rm E} {\cal E}_{\rm HB}$, given by
\begin{eqnarray}
\delta_{\rm E} {\cal E}_{\rm HB}
=\delta E_n \frac{\partial}{\partial {\cal E}_n}{\cal E}_{\rm HB} \ ,
\end{eqnarray}
by numerical
differentiation of ${\cal E}_{\rm HB}$. We employ an effective infinitesimal
displacement of the energy variable obtained by making the replacement of
$E_n(Z\alpha)$ by $E_n [(Z\pm \epsilon )\alpha]$, with
$\epsilon=0.002$.  The symmetric derivative formula with this displacement
has an uncertainty of a few parts in $10^6$. 
One expects that numerical
integration errors, as discussed in \cite{iam98}, are slowly varying
functions of $Z$ and largely cancel in forming the derivative, 
leading to an accurate result for the derivative.  
Evidently, however, the error in terminating the sum over $\kappa$
does not cancel as strongly, which
may lead to significant uncertainties for some $Z$ and 
$n\ell$.  We have thus found it preferable 
to differentiate directly the running
term in the sum over angular momenta in Eq.~(\ref{eq:hb}).  To study
this effect, we compare numerically $\frac{\partial}{\partial E_n }S_A(r,y,u)$
evaluated by differentiation of the sum
\begin{eqnarray}
  \label{eq:dirsum}
\frac{\partial}{\partial E_n }S_A(r,y,u)&=&\frac{\partial}{\partial E_n } \sum_{|\kappa|= 1}^{\infty}
\bigg[K_{\phi,\kappa}(r y,y,iu)   \nonumber \\
&&-
K^{(0,0)}_{\phi,\kappa}(r y,y,iu)-K^{(0,1)}_{\phi,\kappa}(r y,y,iu) \nonumber \\
&& -K^{(1,0)}_{\phi,\kappa}(r y,y,iu)\bigg]  
\end{eqnarray}
to a sum of derivatives (term-by-term differentiation)
\begin{eqnarray}
  \label{eq:tbytsum}
\frac{\partial}{\partial E_n }S_A(r,y,u)&=& \sum_{|\kappa|= 1}^{\infty}\frac{\partial}{\partial E_n }
\bigg[K_{\phi,\kappa}(r y,y,iu) 
  \nonumber \\
&&-K^{(0,0)}_{\phi,\kappa}(r y,y,iu)-K^{(0,1)}_{\phi,\kappa}(r y,y,iu) \nonumber \\
&& - K^{(1,0)}_{\phi,\kappa}(r y,y,iu)\bigg]  
\end{eqnarray}
where $u=(1/2)(1/t - t)$ and $K^{(i,j)}_{\phi,\kappa}$ are defined in
Eqs. (\ref{eq:Kkdf}) to (\ref{eq:K10kdf}). The summation is terminated
when the remainder estimate as described in Ref.~\cite{iam98} is
smaller than a predefined cutoff value. Although the two methods of
calculation converge to the same value for a very small cutoff
($10^{-10}$), the results with a larger cutoff can disagree by a few
parts in $10^{-4}$. Since very small cutoff values would lead to other
problems when r is close to 1, we evaluate the
energy derivative by the term-by-term method. Other illustrations of
this problem are discussed in Sec. \ref{sec:coulomb-tests}.

\section{Analytic Terms}
\label{sec:an_sub}
In this section we evaluate analytically the terms that are 
subtracted in the numerical calculation described in 
Sec.~\ref{sec:high_e} for each of the three
contributions to the screened self-energy.  
In Refs.~\cite{iam92,iam98}, with this application in mind, 
we were careful not to obscure the origins of $V$ and $E_n$ 
through the use of the Dirac equation.  The 
individual terms corresponding to the three 
diagrams of Fig.~\ref{fig:diag_s} are thus
obtained by direct differentiation.

\subsection{Wave-function correction}
\label{sec:wf_sub}
The analytic portion of the wave-function correction follows 
from Refs.~\cite{iam92,iam98}. In each of the terms $E_{\rm H}^{(i,j)}$,
we calculate the variation with respect to a change in the wave function
based on the explicit dependence on the wave function, and obtain
\widetext
\begin{eqnarray}\delta_{\phi} {\cal E}_{\rm H}^{(0,0)}=\frac{2 \alpha}{\pi} && 
\left< \phi_n | \beta |\delta \phi_n \right> \left[\ln({\mit \Lambda}^2) -1 + 
\frac{1-E_n^2}{ E_n^2}\ln\left(1+E_n^2\right) + {\cal O}({\mit \Lambda}^{-1})\right] \ ,
\end{eqnarray} 
\noindent
\begin{equation} 
\delta_{\phi}  {\cal E}_{\rm H}^{(0,1)}=\frac{2 \alpha}{ \pi}
\left< \phi_n | \bbox \alpha \cdot \bbox p \ |\delta \phi_n \right> 
\biggl[\frac{1}{4}\ln({\mit \Lambda}^2)
- \frac{6-3E_n^2+7E_n^4}{ 24E_n^2(1+E_n^2)} 
+ \frac{1-E_n^4}{ 4E_n^4} \ln\left(1+E_n^2\right)
 + {\cal O}({\mit \Lambda}^{-1})\biggr] \ ,
\label{eq:wf00}
\end{equation}
and
\begin{eqnarray} 
\delta_{\phi}  {\cal E}_{\rm H}^{(1,0)}& =& \frac{2 \alpha}{ \pi}  \left\{
\left< \phi_n | V | \delta \phi_n \right> \left[\frac{1}{4}\ln({\mit \Lambda}^2) 
+ \frac{6-E_n^2}{ 8E_n^2} 
- \frac{3+E_n^4}{ 4E_n^4}\ln\left(1+E_n^2\right)\right] \right. 
\nonumber \\
&-&\left. \left< \phi_n | \beta V | \delta \phi_n \right>
\left[ \frac{2}{E_n}-\frac{2}{ E_n^3} \ln\left(1+E_n^2\right) 
\right] + {\cal O}({\mit \Lambda}^{-1})\right\} \ .
\label{eq:wf01}
\end{eqnarray} 
\narrowtext\noindent
In analogy with \cite{iam92,iam98}, we define
\begin{eqnarray} 
\delta_{\phi} {\cal E}_{\rm HA} 
&=& \lim_{{\mit \Lambda}\to \infty}\bigg[\delta_{\phi}  {\cal E}_{\rm H}^{(0,0)} 
+ \delta_{\phi}  {\cal E}_{\rm H}^{(0,1)} 
\nonumber\\ &&
+\delta_{\phi}  {\cal E}_{\rm H}^{(1,0)} 
- 2 \, \delta m ({\mit \Lambda}) 
\left< \phi_n | \beta | \delta \phi_n \right>\bigg] \ ,
\label{eq:wfrenorm}   
 \end{eqnarray} 
\noindent where the last term is the renormalization term and 
\begin{equation} 
\delta m({\mit \Lambda})=\frac{\alpha}{\pi}\left[\frac{3}{4} 
\ln({\mit \Lambda}^2)+\frac{3}{8}\right] \ . 
\end{equation} \noindent 

If we combine the coefficients of 
$\ln ( {\mit \Lambda}^2)$ in Eq.~(\ref{eq:wfrenorm}), we obtain
\begin{equation}
\frac{\alpha}{ 2 \pi}\left< \phi_n | 
\beta + \bbox \alpha \cdot \bbox p+V  |\delta \phi_n \right>
=\frac{\alpha}{ 2 \pi} E_n \left< \phi_n |\delta \phi_n \right>=0
\end{equation}
from the differential equation for $\phi_n$ and the fact that the
first-order correction to the wave function is orthogonal
to the unperturbed wave function. 
Hence, the wave function correction is separately finite.

\subsection{Energy correction}
\label{sec:en_sub}
This correction is obtained by differentiating all terms with respect to ${\cal E}_n$.
The three terms are
\widetext
\begin{eqnarray}
\delta_{\rm E}  {\cal E}_{\rm H}^{(0,0)}=\frac{\alpha}{\pi}\delta E_n &&\left[
\left<\phi_n|\beta|\phi_n\right>
\left(\frac{2}{E_n}\, \frac{1- E_n^2}{1+E_n^2}
-\frac{2}{ E_n^3} \ln\left(1+E_n^2\right) 
\right) \right.\nonumber\\ &&
\left.-\frac{1}{4}\ln({\mit \Lambda}^2)-\frac{6-E_n^2}{ 8E_n^2} 
+ \frac{3+E_n^4}{ 4E_n^4}\ln\left(1+E_n^2\right)
+ {\cal O}({\mit \Lambda}^{-1})\right] \ ,
\end{eqnarray} 
\begin{eqnarray} 
\delta_{\rm E}  {\cal E}_{\rm H}^{(0,1)}=\frac{\alpha}{\pi}\delta E_n
\left<\phi_n | \bbox \alpha \cdot \bbox p | \phi_n \right>&& \biggl[
\frac{6+9E_n^2-8E_n^4-3E_n^6}{6E_n^3\left(1+E_n^2\right)^2}
 - \frac{1}{E_n^5} \log (1 + {E_n^2}) 
 + {\cal O}({\mit \Lambda}^{-1})\biggr] \ , 
\end{eqnarray}
\noindent and 
\begin{eqnarray}
\delta_{\rm E}  {\cal E}_{\rm H}^{(1,0)}=\frac{\alpha}{\pi} \delta E_n &&\left\{
\left<\phi_n|V|\phi_n\right> \left[
-\frac{6+3E_n^2+E_n^4}{2E_n^3(1+E_n^2)}
 +  \frac{3}{E_n^5} \log (1 + {E_n^2})
\right] \right. \nonumber \\
+\left<\phi_n|\beta V|\phi_n\right> &&  \left. \left[ 
    \frac{6+2E_n^2}{E_n^2(1+E_n^2)}
    -\frac{6}{E_n^4}\log( 1 + E_n^2 )
\right] + {\cal O}({\mit \Lambda}^{-1})\right\} \ .  
\end{eqnarray}
\narrowtext \noindent
The total is 
\begin{eqnarray} 
\delta_{\rm E} {\cal E}_{\rm HA} &=&\delta_{\rm E}  {\cal E}_{\rm H}^{(0,0)} 
+ \delta_{\rm E}  {\cal E}_{\rm H}^{(0,1)} +\delta_{\rm E}  {\cal E}_{\rm H}^{(1,0)} \ ,
\label{eq:erenorm}   
\end{eqnarray} 
which contains a divergent term given by
\begin{equation} 
\delta_{\rm E} {\cal E}_{\rm HA} =  
-\frac{\alpha \,\delta E_n}{4\pi}\ln({\mit \Lambda}^2) + {\cal O}(1) \  .
\label{eq:erenorml}   
\end{equation}

\subsection{Green's function correction}
\label{sec:gre_sub}

This term arises entirely from $ {\cal E}_{\rm 
H}^{(1,0)}$, which is linear in the potential in the 
Green's function.  Taking into account the fact that
$\left<\phi_n|\delta V|\phi_n\right> = \delta E_n$, we obtain
\widetext
\begin{eqnarray}
\delta_{\rm G} {\cal E}_{\rm HA} =
\delta_{\rm G}  {\cal E}_{\rm H}^{(1,0)}&=&\frac{\alpha}{\pi}  \bigg\{
\delta E_n  \bigg[\frac{1}{4}\ln({\mit \Lambda}^2) 
+\frac{6-E_n^2}{ 8E_n^2} 
- \frac{3+E_n^4}{ 4E_n^4}\ln\left(1+E_n^2\right)\bigg]  \nonumber \\ &&
-  \left<\phi_n|\beta\delta V|\phi_n\right>  
\left[\frac{2}{E_n}-\frac{2}{ E_n^3} \ln\left(1+E_n^2\right)\right]  
+ {\cal O}({\mit \Lambda}^{-1})\bigg\}\bigg] \ .   
\end{eqnarray}
\narrowtext\noindent
This expression also contains a divergent contribution
\begin{eqnarray} 
\delta_{\rm G} {\cal E}_{\rm HA} =
\frac{\alpha\,\delta E_n }{4\pi}\ln({\mit \Lambda}^2) + {\cal O}(1) \ ,
\label{eq:grenorm}   
 \end{eqnarray} 
which cancels the corresponding term in Eq.~(\ref{eq:erenorml}).
The total expression 
\begin{eqnarray}
\delta E_{\rm HA} = 
\delta_{\phi} {\cal E}_{\rm HA}
+ \delta_{\rm E} {\cal E}_{\rm HA}
+ \delta_{\rm G} {\cal E}_{\rm HA}
\end{eqnarray}
is thus finite as expected.

\section{Results and discussion}
\label{sec:results}
\subsection{Coulomb tests}
\label{sec:coulomb-tests}
\begin{figure*}[tbp]
        \centering
        \includegraphics{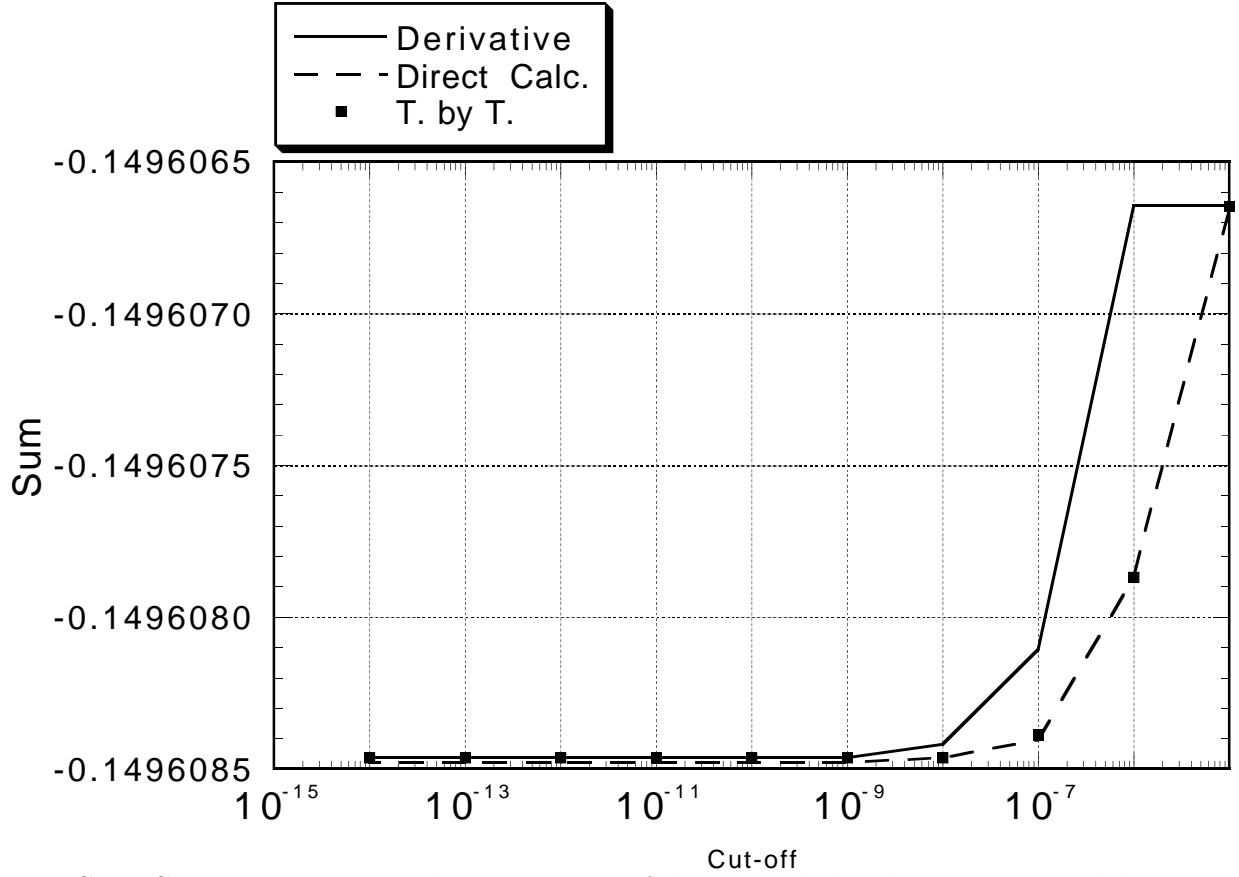}
        \caption{Comparison between the convergence of the sum
        calculated using numerical derivation of the unperturbed
        self-energy sum as in Eq.~(\protect \ref{eq:hb}) to the direct
        evaluation using perturbed wave function and reduced Green's
        function as defined in Eq.~(\protect\ref{eq:dirsum}), and to
        an evaluation in which terms in Eqs.~(\ref{eq:Kk}) to
        (\ref{eq:K10k}) are derived for each $\kappa$ value in the sum
        (T.-by-t deriv.). Evaluation is done for a $2p_{3/2}$ state and $r = 0.9992$, $y=1$,
        $t=0.99$, $Z=20$.}
        \label{fig:sumcomp}
\end{figure*}

In order to check the accuracy of the numerical calculation, the
equations derived in the previous sections, and the parallelized code,
we compare the calculations obtained by numerical differentiation of the
one-electron self-energy function $F(Z\alpha)$ to the results of the
method presented in this paper using $\delta V
(r)=-\epsilon\,\alpha/r$ as a perturbing potential.  This method, proposed
in Ref.~\cite{iam91}, is very efficient, as each individual
contribution to the screened self-energy can be checked independently.
The overall agreement between the results of these two methods of
calculation is good, although
differences between some contributions can be several times the
combined uncertainties based only on the apparent convergence of the
numerical integration. These additional errors come from the
numerical problems described in Sec.~\ref{sec:e_high}. An
illustrative example is the wave function
correction. One can compare the term-by term derivative with respect
to $Z$ in the wave function to the derivative obtained from
two converged sums, as is done in Eqs.~(\ref{eq:dirsum}) and 
(\ref{eq:tbytsum}).  There is also an independent calculation
based on the first-order correction to the wave function from Eq.
(\ref{eq:dwf}). The evolution of this sum for smaller and smaller
values of the cutoff error are displayed in Fig.~\ref{fig:sumcomp}. 
One can
see that although the two calculations converge to the same limit when
the cutoff is as small as $10^{-15}$, the results follow very different
paths. Only the term-by-term differentiation method follows the 
result obtained from evaluation of the first-order correction 
to the wave function independently of the cutoff. 
This constitutes a very demanding test of
our numerical evaluation of the first-order correction to the wave
function.  The difference between the two calculations is never
smaller than $8\times 10^{8}$, which is the error from numerical 
uncertainties of the full Green's
function and the error in the numerical derivative.  

To improve the numerical precision, we have employed a subdivision
of the integration over $r$ into regions with $0<r<0.4$ and
with $0.4<r<1$, as described in Ref.~\cite{iam98}.
This division provides an accurate evaluation that does not 
require functional evaluations with values of
$r$ too close to 1. In this way we where able to obtain an
accurate comparison of all contributions in the high-energy part.

A few such problems remain in the test calculations in the low-energy
part at low $Z$.  We did not attempt to improve the accuracy, because
we have enough accurate cases to check the code and numerical
procedures.  From the tests we have performed, it is clear that the
calculation based on numerical differentiation is the less accurate.
However since numerical differentiation is used in the final result to
obtain the reducible correction we have increased the total
uncertainty of both the pure Coulomb test and spherically-averaged
potential of the next section accordingly.

\subsection{Results with spherically-averaged one-electron potential}
The calculations of interest for physical applications are based on
realistic potentials obtained from the spherically-averaged
potential of Eq.~(\ref{eq:mods}) of Appendix~\ref{app:modpot}.  All
results presented here are given in terms of the scaled function defined by
\begin{equation}
{\cal E}_{n\ell j,n^{\prime}\ell^{\prime} j^{\prime}}={\alpha^2
(Z\alpha)^3 \over \pi n^3}F_{n\ell j,n^{\prime}\ell^{\prime} j^{\prime}}(Z\alpha)
\,mc^2
\label{eq:scaled} 
\end{equation}

\begin{figure*}[tbp]
        \centering
        \includegraphics[width=14cm]{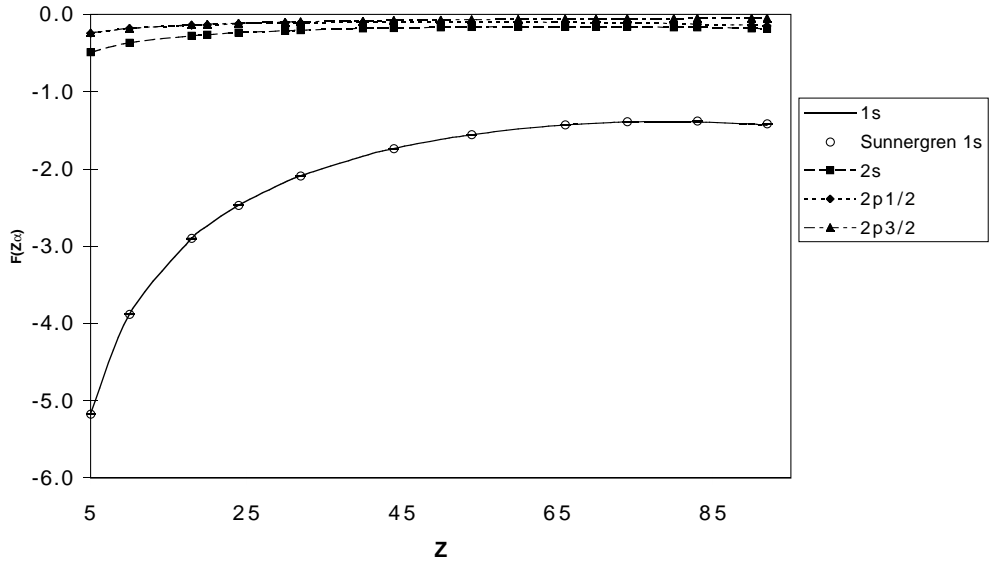}
        \caption{Screening of a $1s$ electron by a $1s$, $2s$, $2p_{1/2}$ or $2p_{3/2}$ electron.
          Sunnergren: Correction to the Coulomb part of the
          electron-electron interaction from
          Ref.~\protect\cite{sun98}, improved calculation following
          Ref.~\protect\cite{pssl96}.}
        \label{fig:1s1s}
\end{figure*}

\begin{figure*}[tbp]
        \centering
        \includegraphics[width=14cm]{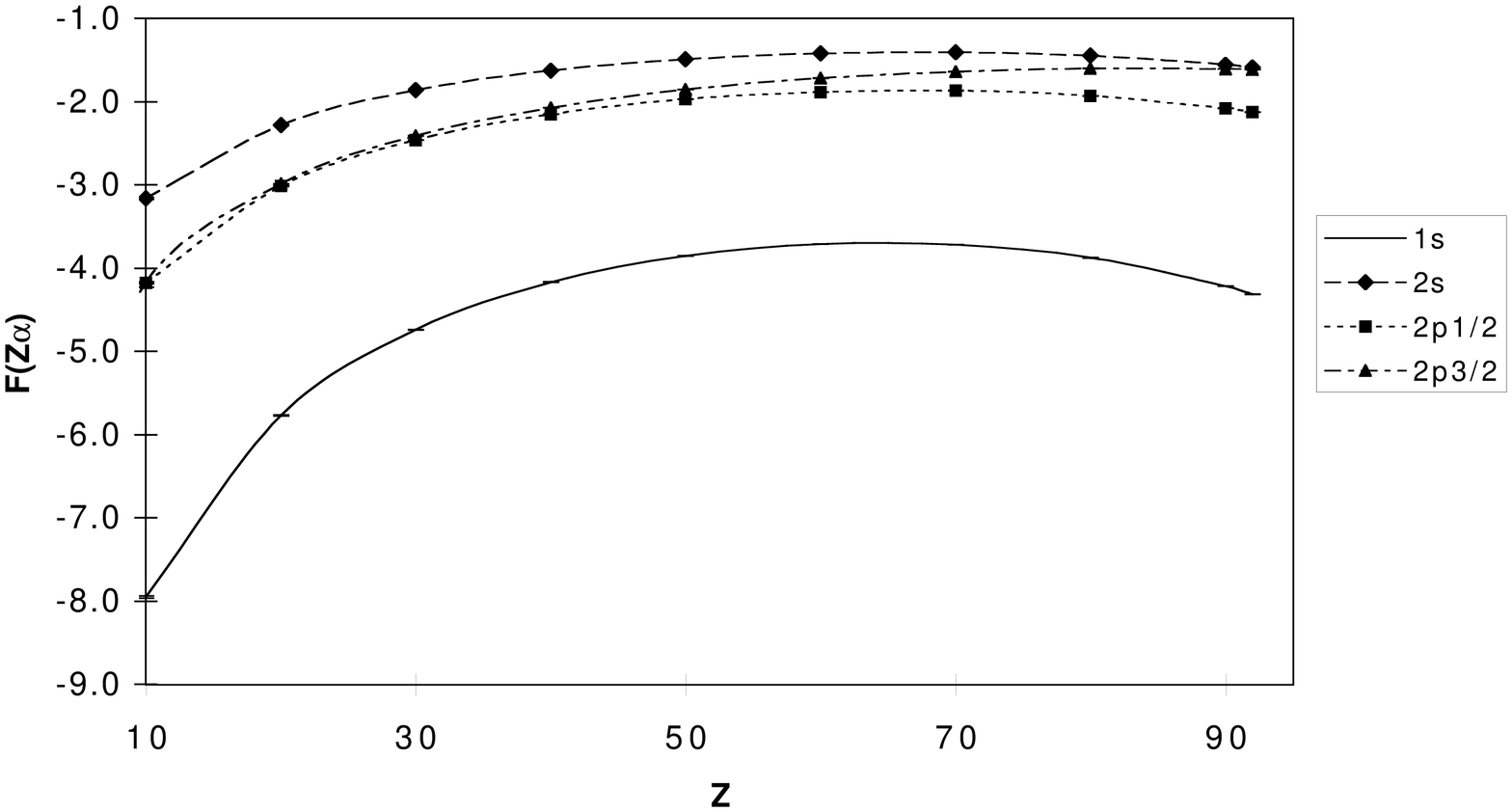}
        \caption{Screening of a $2s$ electron by a $1s$, $2s$, $2p_{1/2}$ or $2p_{3/2}$ electron.}
        \label{fig:2s1s}
\end{figure*}

\begin{figure*}[tbp]
        \centering
        \includegraphics[width=14cm]{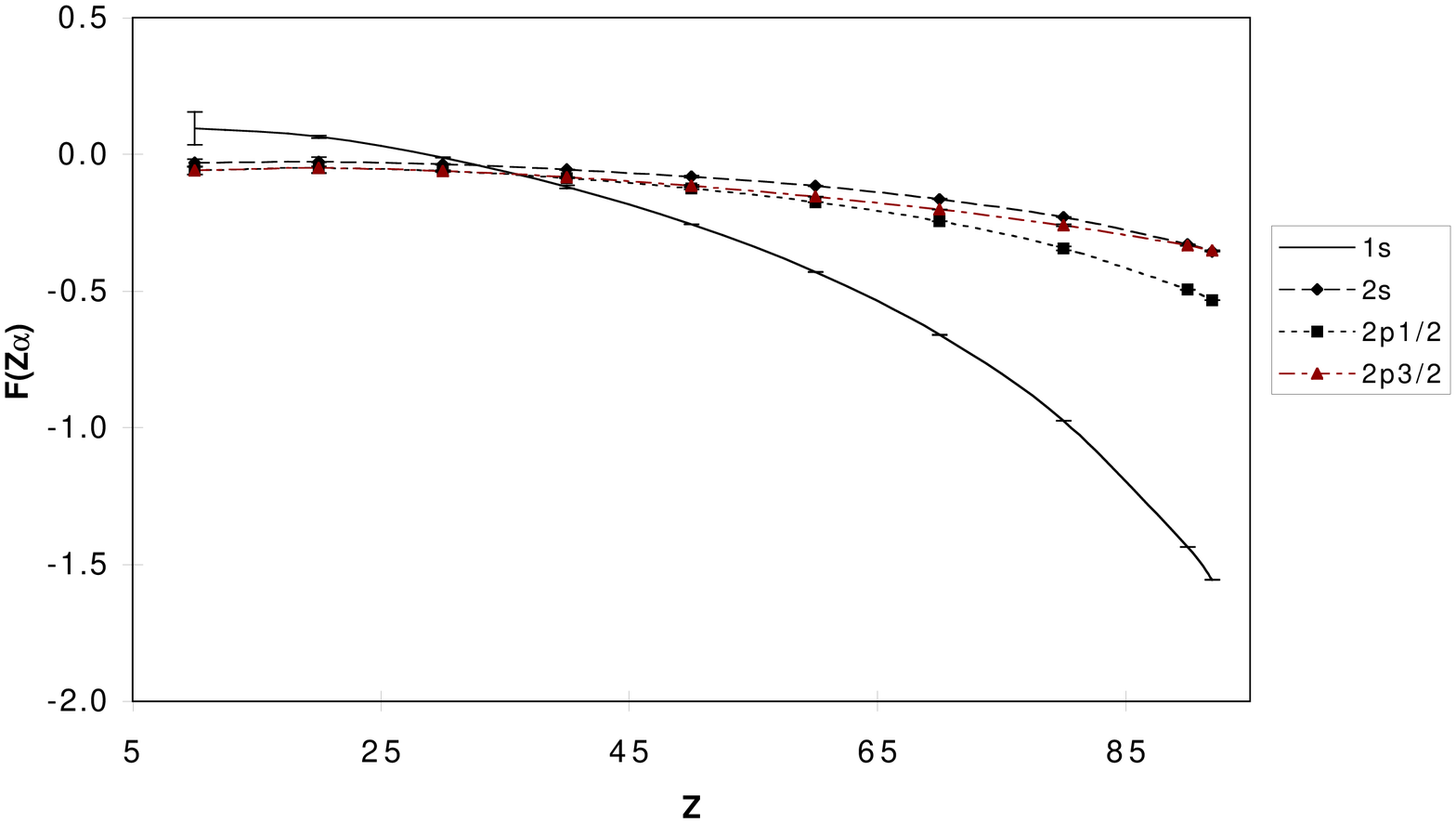}
        \caption{Screening of a $2p_{1/2}$ electron by a $1s$, $2s$, $2p_{1/2}$ or $2p_{3/2}$ electron.}
        \label{fig:2p12p1}
\end{figure*}

\begin{figure*}[tbp]
        \centering
        \includegraphics[width=14cm]{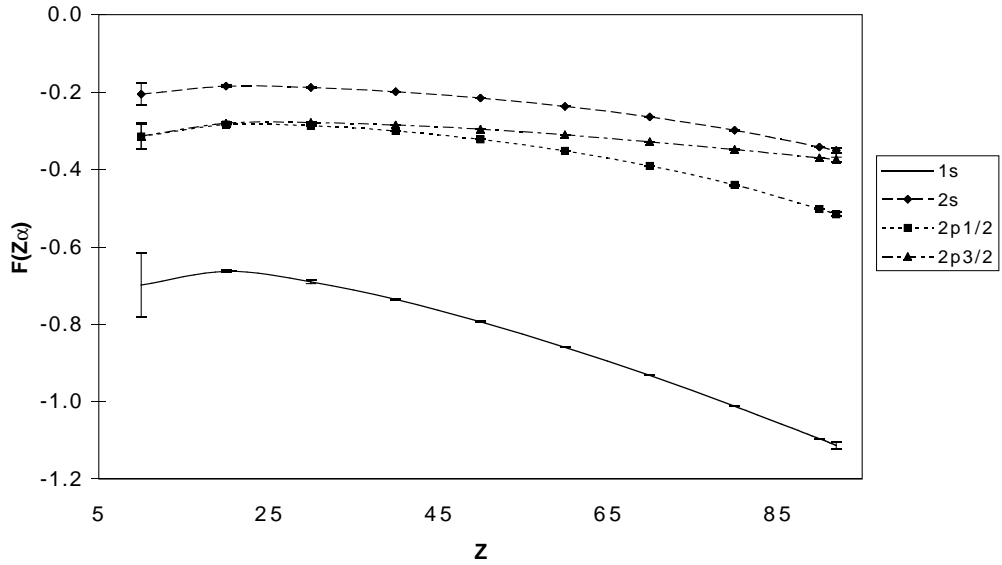}
        \caption{Screening of a $2p_{3/2}$ electron by a $1s$, $2s$, $2p_{1/2}$ or $2p_{3/2}$ electron.
 The large error bar for $Z=10$ are only due to the low-energy part of the vertex correction}
        \label{fig:2p32p1}
\end{figure*}

All 16 possible total scaled functions $F(Z\alpha)$ for the self-energy
screening of $n\ell$ electron by a $n^{\prime}\ell^{\prime}$ electron,
$1\leq n,\, n^{\prime}\leq 2$, $0\leq \ell,\, \ell^{\prime}\leq 1$ are
given in Table \ref{tbl:totcont} and in Figs.~\ref{fig:1s1s} to
\ref{fig:2p32p1}. It can be seen that the uncertainty at low $Z$ can be
as high as 30\% for the screening of $2p$ electrons at $Z=10$, or as
low as $10^{-6}$.  These functions can be used to evaluate the
self-energy screening correction to any atom with two to ten
electrons, in a shell $n\leq 2$. As an example we treat the case of
lithiumlike uranium. With the results presented here we can compute
the self-energy screening correction for all three states $1s^2 2s$, $1s^2
2p_{1/2}$ and $1s^2 2p_{3/2}$.  A first approximation is obtained
for transition energies by neglecting the core relaxation. The
self-energy screening correction to $2p_{j}\to 2s$ transition energy
is evaluated using Eq.~(\ref{eq:scaled}) as $\Delta {\cal E}_{j}= 2
{\cal E}_{2p_j, 1s}-2 {\cal E}_{2s, 1s}$, where the factor of two
accounts for the fact that there are two $1s$ electrons screening
$n=2$ electron.  A better approximation, which takes into account the
relaxation of the core electrons, and provides a value for the
total binding energy is given by ${\cal E}_{1s^2 2s} = 2({\cal E}_{2s,
  1s}+{\cal E}_{1s, 2s}+{\cal E}_{1s, 1s})$ and ${\cal E}_{1s^2 2p_j}
= 2({\cal E}_{2p_j, 1s}+{\cal E}_{1s, 2p_j}+{\cal E}_{1s, 1s})$.  The
results of these calculations are presented in Table \ref{tbl:lilikeu}
together with all other calculations known to date.

It should be noted that the present method is equivalent to
the Coulomb approximation in certain cases.
If one considers only the {\it Coulomb contribution} in the
interaction between the two electrons in Fig.~\ref{fig:diagred}, the
present method provides the equivalent contribution
to the ground state of two-electron ions. This follows from the
fact that between $s$ states, only the monopole part of the
$1/r_{12}$ operator contributes. This radial contribution of the
monopole part is exactly given by the potential in
Eq.~(\ref{eq:perportexp}). Moreover the
retarded part of the Coulomb interaction vanishes in this case.
Finally, the exchange correction only involves
the spin of the two electrons and corresponds to a multiplication
of the function $F_{1s,1s}(Z\alpha)$ by two.

The above arguments can be extended for all cases where an electron of
arbitrary quantum number interacts with a $s$ electron. The $s$
electron couples only to the monopole term in the angular expansion of the
Coulomb interaction. If, however, the electrons are not identical,
then there is an exchange term with additional
multipole terms. In this case, 
the retardation contribution to the Coulomb
interaction is also non-zero.
Obviously in the relativistic case the magnetic part of the
electron-electron interaction should be considered.

\begin{figure*}[tbp]
        \centering
        \includegraphics[width=14cm]{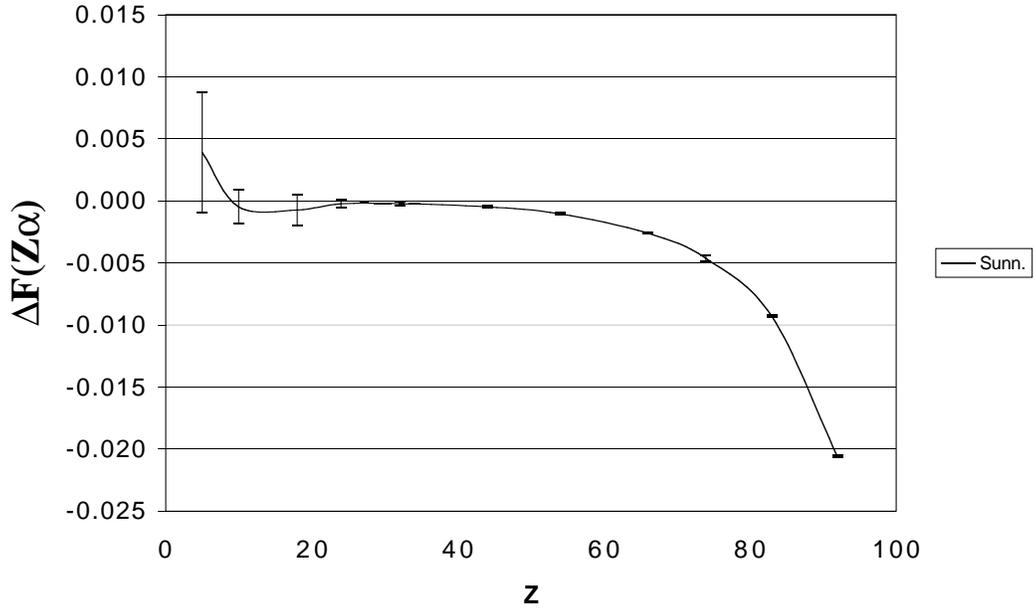}
        \caption{Comparison between calculation using the partial-wave renormalization method \protect\cite{pssl96}
          and this work. More accurate numerical values from
          Ref.~\protect\cite{sun98} are used.}
        \label{fig:compsun}
\end{figure*}

Because of these considerations we can compare our results to the {\it
  Coulomb} part of the calculation done by the G\"{o}teborg Group. In
Fig.~\ref{fig:1s1s} we plot also the function $F(Z\alpha)$ from 
Refs.~\cite{pssl96,sun98}.  The difference between the two calculations is
displayed on Fig.~\ref{fig:compsun}.
The agreement is very good for
medium-$Z$, while the difference between the two calculations
increases with increasing $Z$, 
which is due to the inclusion of finite nuclear
size in Refs. \cite{pssl96,sun98}, while the present results are 
for a point nucleus. This comparison thus provides the
finite nuclear size effect on the two-electron self-energy.  The difference
at low $Z$ (5 and 10) are due to numerical inaccuracies.  Since no
uncertainties are given in Ref.~\cite{sun98}, which contains more
accurate values than Ref.~\cite{pssl96},
we assume an uncertainty of 1 in the last digit (note that
$F(Z\alpha)$ in Ref.~\cite{sun98} is two times ours.)

\begin{figure*}[tbp]
        \centering
        \includegraphics[width=14cm]{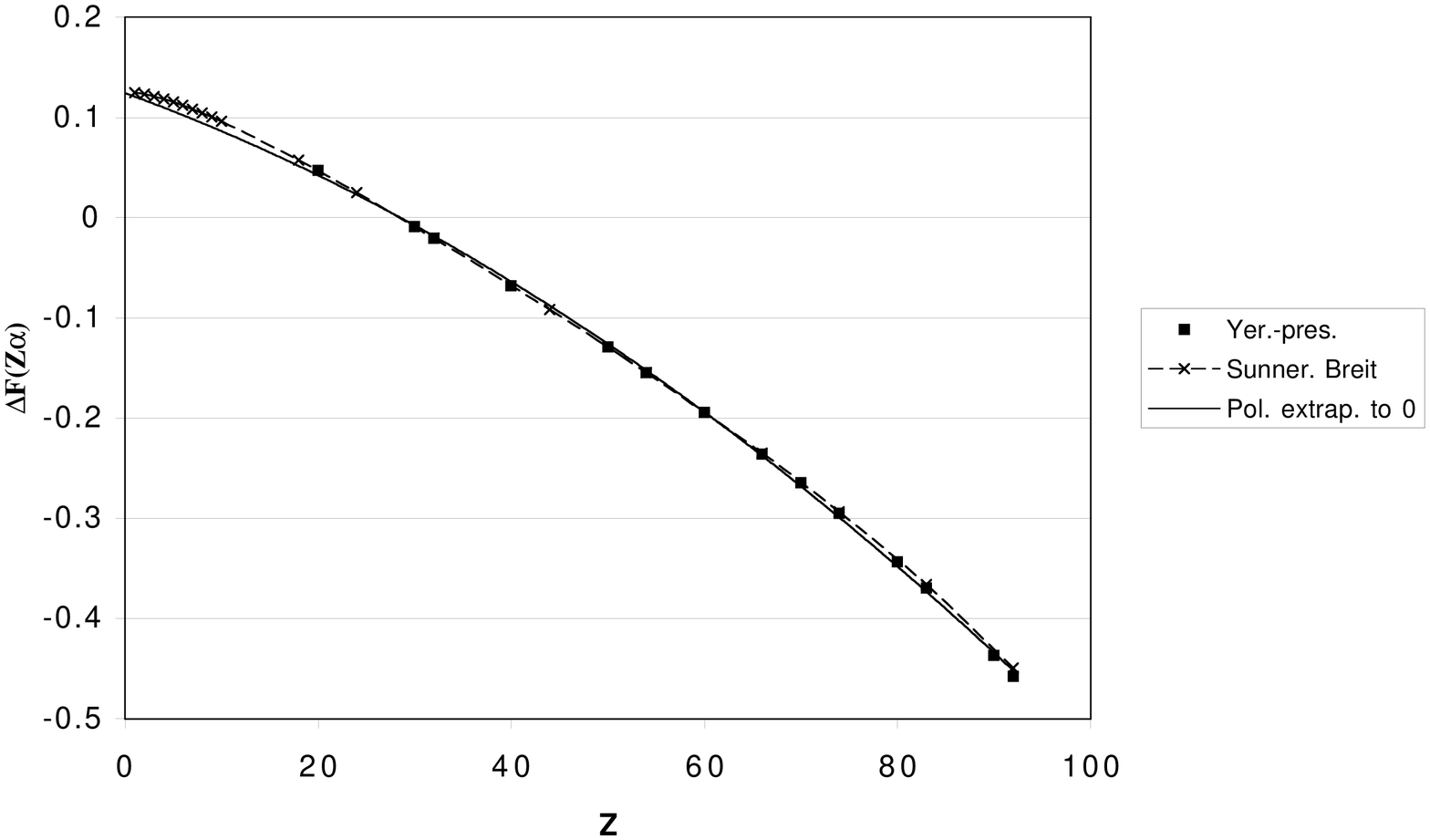}
        \caption{Comparison between calculation for point nucleus 
          (including Coulomb and magnetic part of the
          electron-electron interaction) from  
          Ref.~\protect\cite{yas97} and the Coulomb value from this 
          work. The solid line represents a second-order
          polynomial fit to the difference and extrapolates to $-0.124$ for $Z=0$, as expected
          from Eq.~(\protect \ref{eq:fmaglz}). Also plotted are the Breit 
          values from Ref.~\protect\cite{sun98}, which are in reasonable agreement 
          with the extrapolation. 
\label{fig:yas97}}
\end{figure*}

The low-$Z$ behavior of the self-energy correction for to the Coulomb
interaction to $F_{1s,1s}(Z\alpha)$ is known from the
work of Araki \cite{ara57} and Sucher \cite{suc58}, and $1/Z$ 
expansions from Drake \cite{dra96} to be
\begin{eqnarray}
  F^{\rm{Coul}}_{1s,1s}(Z\alpha)&=&\frac{1}{2}\bigg[2.588819\nonumber \\
&& +\left(\frac{7}{2}-2\log 2\right) 
\log\left(2\left(Z\alpha\right)^2\right)\bigg],
\label{eq:fcoullz}
\end{eqnarray}
while the magnetic part is
\begin{equation}
  F^{\rm{mag}}_{1s,1s}(Z\alpha)=\frac{1}{8}
\label{eq:fmaglz}
\end{equation}
We cannot directly compare our value for $F_{1s,1s}(Z\alpha)$ with
those of Yerokhin et al. \cite{yas95,yas97}, because their results
also include the magnetic and retardation contribution to the self
energy and our model does not. This correction contributes even at
very low $Z$ since it contains the free-electron anomalous magnetic
moment from the vertex correction. The difference between the total
contribution (for point nucleus) from Ref.~\cite{yas97} and the
present work is plotted on Fig.~\ref{fig:yas97} for $20 \ge Z$.
Evidently, the magnetic interaction contribution to the two-electron
self-energy is much larger than the contribution of the finite nuclear
size. A simple fit of the difference between the present result and
the one in Ref.~\cite{yas97} with a second-order polynomial yields
$0.124$ for the contribution of the anomalous magnetic moment, in good
agreement with the value in Eq. (\ref{eq:fmaglz}).  From the figure,
it is evident that for $Z$ as low as 5 the higher-order terms still
make a significant contribution.  On the same figure we also plot  the
Breit contribution from Ref.~\cite{sun98}, which is in agreement with
the difference between Yerokhin et al. and the present work for $Z \ge
20$, and matches reasonably well the extrapolated values even down to $Z=1$.

\section{Conclusion}
\label{sec:concl}

In this paper we describe a method of approximately evaluating
two-electron radiative corrections that can easily be generalized to
the direct evaluation of the correction represented by the diagrams in
Fig.~\ref{fig:diagred}. Accuracy and correctness of the method and
programs is assessed by extensive comparisons with numerical
derivatives of well-known one-electron self-energy results for a
Coulomb perturbation.  It is demonstrated 
that the method can work down to
$Z=5$ in some cases with reasonable accuracy.  With the use of a more
accurate Green's function evaluation and convergence acceleration
techniques, following Refs.~\cite{jms99,jmsw99}, it is likely
that calculation can be performed for He. 
The results presented in the present
paper also provides approximate self-energy screening
corrections in any ion with less than 10 electrons, thus providing a
valuable, QED-based replacement for methods based on the Welton
approximation \cite{igd87,iad90} or other, less efficient, screening
schemes as used in atomic structure codes.  It is also
equivalent to the direct Coulomb contribution for some states of
helium-like ions.

This method could also be used with
numerical Dirac-Fock potentials and
wave functions from one of the codes in Refs. \cite{des93,dgjpp89}.
Preliminary tests show that good numerical accuracy can be achieved.
Such an approach would provide more accurate self-energy screening
corrections for the outer shells of very heavy transuranic elements or
for inner hole binding energies \cite{ibl98}.

\acknowledgments 

The numerical calculations presented here have been made possible by a 
generous computer time allocation on the IBM SP2 at the Centre 
National Universitaire Sud de Calcul (Montpellier, France).  Some of 
the calculations and computer program development were
done on the NIST SP2.

\appendix

\narrowtext

\section{Numerical test}
\label{app:numtest}
As a consistency check on the computer code, we carry out a test 
calculation in which the correction terms are generated by numerical 
differentiation of the unperturbed Coulomb self energy with respect to 
the nuclear charge $Z$.  This should give the same result as the 
screening calculation where both the unperturbed potential and the 
perturbing potential are the Coulomb potential, with an appropriate 
normalization factor.

In other words, we consider the potential
\begin{eqnarray}
V(x) = -\frac{(Z+\Delta Z)\alpha}{ x}
\end{eqnarray}
and let
\begin{eqnarray}
V(x) = V_0(x) + \delta V(x)
\end{eqnarray}
where
\begin{eqnarray}
V_0(x) &=& -\frac{Z\alpha}{ x}
\nonumber\\
\delta V(x) &=& - \frac{\Delta Z \alpha}{ x}
\end{eqnarray}
If the level shift  $E(Z)$ is known as a function
of $Z$ for the Coulomb potential, 
then the exact correction due to $\delta V(x)$ is $E(Z+\Delta Z) - E(Z)$
and the first-order correction in $\Delta Z$ is 
\begin{eqnarray}
\delta E(Z) &=& \Delta Z \lim_{\delta Z \rightarrow 0}\frac{E(Z+\delta Z) - E(Z)}
{ \delta Z}
\nonumber\\
&=&\Delta Z \frac{\partial}{\partial Z} E(Z) \equiv \Delta Z E^\prime(Z)
\label{eq:ddef}
\end{eqnarray}
Thus the first-order perturbation due to the potential $\delta V(x)$, 
with unit charge shift $\Delta Z = 1$ should be exactly equal to the 
derivative with respect to $Z$ of the Coulomb level shift 
$E^\prime(Z)$.


\section{Origin expansion of the first-order correction to the wave function}
\label{app:orexpwf1}
This origin expansion is made with the use of the differential equation for the 
first-order correction to the wave function in Eq.~(\ref{eq:fodifdirac}).
However this expansion has a different form depending on whether one uses
a Coulomb perturbing potential or a potential created by an other electron.

\subsection{Coulomb perturbation potential}
\label{sec:coulpot}
In the case of a Coulomb perturbing potential, the correction to the
wave function must have a logarithmic contribution, so as to cancel
the $\alpha/x$ term in the lowest order of the development.
We write  
\begin{eqnarray}
 \delta f_i(x)&=&\alpha x^{\omega}\bigg[\sum_{j=0}^{\infty} \zeta_i^{(j)} \frac{x^j}{j!}
+\log(x) \sum_{j=0}^{\infty} \lambda_i^{(j)} \frac{x^j}{j!}\bigg],
\label{eq:couldwf1}
\end{eqnarray}
and replace in Eq.~(\ref{eq:fodifdirac}), together with a series
expansion of the unperturbed wave function, which behaves as $\omega$
near the origin. We then extract coefficients of $x^j \log (x)$ and of
$x^j$ and solve for the $\zeta^{(j)}_i$ and $ \lambda^{(j)}_i$ coefficients.  The
coefficient of the $\log (x)/x$ term for an unperturbed wave function
of angular symmetry $\kappa_n$ is
\begin{equation}
  \label{eq:dwflogx-1}
\left\{
  \begin{array}{l}
(-\kappa_n-\lambda)\lambda_0^{(1)} +\gamma\lambda_0^{(2)}=0\\
-\gamma \lambda_0^{(1)}+(\kappa_n-\lambda) \lambda_0^{(2)}=0
  \end{array}
\right.
\end{equation}
where the unperturbed wave function origin behavior is given by $\omega=
\lambda-1$, $\lambda=\sqrt{\kappa_n^2-\gamma^2}$, $\gamma=Z\alpha$. The determinant of this equation
is zero and thus we can write 
\begin{equation}
 \label{eq:lam0}
 \lambda_0^{(1)}=\frac{\kappa_n-\lambda}{\gamma}\lambda_0^{(2)}.
\end{equation}
The general equation for the term of order $i$ is
\begin{equation}
  \label{eq:dwflogx}
\left\{
  \begin{array}{l}
(-\kappa_n-\lambda-i)\lambda_i^{(1)} +\gamma\lambda_i^{(2)}=(1+E_n)\lambda_{i-1}^{(2)}\\
-\gamma \lambda_i^{(1)}+(\kappa_n-\lambda-i) \lambda_i^{(2)}=(1-E_n)\lambda_{i-1}^{(1)}
  \end{array}
\right.
\end{equation}
The determinant of the linear system in Eq.~(\ref{eq:dwflogx}) is
given by $(2\lambda+i)i$ and is nonzero for $i > 0$. By solving order
after order, all higher-order terms can be expressed as a function of
$\lambda_0^{(2)}$, $\gamma$ and $\kappa_n$. The expressions are all
relatively simple since the unperturbed wave function does not have a
logarithmic contribution.  The non-logarithmic terms are obtained by first defining 
a series expansion for the unperturbed wave function
\begin{eqnarray}
 f_i(x)&=&(-1)^{i-1} N_{i} x^{\omega}\bigg[\sum_{j=0}^{\infty} {\varphi}_i^{(j)} \frac{x^j}{j!}
\bigg],
\label{eq:unpertwf}
\end{eqnarray}
 where $i=1$, 2 and $N_{i}$ is a normalization factor.
The equation derived from the term of order $1/x$ is given by
\begin{equation}
  \label{eq:dwfx0}
\left\{
  \begin{array}{l}
(-\kappa_n-\lambda)\zeta_0^{(1)} +\gamma\zeta_0^{(2)}=
N_{2}{\varphi}_2^{(0)}+ \frac{\lambda-\kappa_n}{\gamma}\lambda_0^{(2)}\\
-\gamma \zeta_0^{(1)}+(\kappa_n-\lambda) \zeta_0^{(2)}=N_{1}{\varphi}_1^{(0)} - \lambda_0^{(2)}
\end{array}
\right. ,
\end{equation}
where we have used Eq.~(\ref{eq:lam0}), and which again has a zero
determinant. Explicit expressions of the $n=1$ and $n=2$ wave
functions can be found in Ref.~\cite{moh82,moh74b} (Note that in
Ref.~\cite{moh82} the norm $N_3$ of the $2p_{3/2}$ wave function in
Eq.~(A3) should be
$N_3=\gamma^{5-2\delta'}/[2\Gamma(5-2\delta')]$). Requiring the
compatibility of the two equations enables to calculate $
\lambda_0^{(2)}$ as
\begin{equation}
  \label{eq:dwfzeta}
 \lambda_0^{(2)} =\frac{(\kappa_n+\lambda)N_{1}{\varphi}_1^{(0)}-\gamma N_{2}{\varphi}_2^{(0)}}
{2\lambda}.
\end{equation}
One can then obtain a relation between $\zeta_0^{(2)}$ and $\zeta_0^{(1)}$   using one of the two equations in (\ref{eq:dwfx0})
\begin{equation}
  \label{eq:dwfzeta10}
 \zeta_0^{(1)} =\frac{\gamma^2\zeta_0^{(1)}+(\kappa_n-\lambda)\lambda_2^{(0)}-\gamma N_{2}{\varphi}_2^{(0)}}
{\gamma(\kappa_n+\lambda)}.
\end{equation}

All $\zeta$ coefficients can thus be expressed as a function of
$\zeta_0^{(2)}$. These coefficients must be determined from the
normalization condition of the perturbed wave function.  This obliges
to explicitly write $\zeta_n^{(i)}$, $n>0$, as a function of
$\zeta_0^{(2)}$ rather than keeping them as function of
$\zeta_{n-1}^{(i)}$. The latter expressions are simpler, but the former
are very large.  We use Mathematica to build the equations obeyed by
the $\zeta_n^{(i)}$ an $\lambda_n^{(i)}$ coefficients, evaluate the
explicit expressions of $\zeta_n^{(i)}$, extracting the part which
depends on $\zeta_0^{(i)}$ and the one which doesn't, and generating
FORTRAN code. The code can have hundreds of lines for each piece of
$\zeta_4^{(i)}$.  The final expression of $df$ can finally be recast as
\widetext
\begin{eqnarray}
\label{eq:couldwf1fin}
\left\{  \begin{array}{l}
 \delta f_1(x)=\alpha x^{\omega}\bigg[\zeta_0^{(2)}\sum_{j=0}^{\infty}
 \zeta_{1,a}^{(j)} \frac{x^j}{j!}+\sum_{j=0}^{\infty} \zeta_{1,b}^{(j)} \frac{x^j}{j!}
+\log(x) \sum_{j=0}^{\infty} \lambda_1^{(j)} \frac{x^j}{j!}\bigg]\\
 \delta f_2(x)=\alpha x^{\omega}\bigg[\zeta_0^{(2)}\left(1+
\sum_{j=1}^{\infty} \zeta_{2,a}^{(j)} \frac{x^j}{j!}\right)+
\sum_{j=0}^{\infty} \zeta_{i,b}^{(j)}
+\log(x) \sum_{j=0}^{\infty} \lambda_2^{(j)} \frac{x^j}{j!}\bigg]
  \end{array}
\right.
\end{eqnarray}
\narrowtext
A comparison for a small value of $x$ of the expansion and of the value obtained by 
the use of the reduced Green's function as described in Sec. \ref{sec:numedphi} yield
two values of $\zeta_0^{(2)}$, one for each component of the wave function.
A comparison of the two values provide a good check of the algebra. We compute
$\zeta_i^{(j)}$ and $\lambda_i^{(j)}$ coefficients up to $j=3$. The value of
$\zeta_0^{(2)}$ obtained from each component of the wave function at $x=0.0005$
agree with an accuracy of 13 significant figures, for $n=1$ and $n=2$, 
$\kappa_n=-1$, $1$, $-2$.

\subsection{Electron screening potential}
\label{app:screenpot}
The screening potential described in Eq.~(\ref{eq:mods}) leads to an
origin expansion rather different than the one described in the
preceding section.  In order to evaluate the origin expansion of the
first-order wave function correction we first evaluate the screening
potential expansion. Using the origin expansion of the wave function 
and Eq.~(\ref{eq:mods}), one can easily show that
\begin{equation}
  \label{eq:perportexp}
  \delta V^{(j)}(x)=\delta V^{(j)}(0)+x^{2\lambda_{(j)}} \left(\sum_{n=0}^{\infty}
\delta V^{(j)}_n \frac{x^n}{n!}\right)
\end{equation}
where $\lambda_{(j)}=\sqrt{\kappa_{(j)}^2-\gamma^2}$ (the origin behavior of the screening wave function is $\lambda_{(j)}-1$) and
\begin{equation}
  \label{eq:dv0}
 \delta V^{(j)}(0)= \alpha\int_0^\infty dx x
\left[f_1^2(x) + f_2^2(x)\right].
\end{equation}
The asymptotic expansion is obtained by substituting Eq.~(\ref{eq:perportexp})
in the Poisson equation obeyed by the potential
\begin{equation}
  \label{eq:dvpoiss}
  \frac{d}{dx^2}(x \delta V^{(j)}(x))+\alpha x\left[f_1^2(x) + f_2^2(x)\right]=0
\end{equation}
 and expanding the two radial component of the wave function in powers of $x$.

With such an expansion the shape of the origin expansion of the first
order correction to the wave function is
\begin{eqnarray}
      \delta f_i(x)&=&\alpha x^{\lambda}\bigg[\frac{\zeta_{i}^{(-1)}}{x} +\sum_{j=0}^{\infty} \zeta_i^{(j)} \frac{x^j}{j!}
\nonumber \\ && +x^{2\lambda_{(j)}} \sum_{j=0}^{\infty} \lambda_i^{(j)} \frac{x^j}{j!}\bigg].
\label{eq:elscdwf1}
\end{eqnarray}
We obtain the equation for $\lambda_0^{(j)}$ by looking at the coefficients of
$x^{2\lambda_{(j)}+\lambda}$. We get for the equation of order $1/x$:

\begin{equation}
  \label{eq:dwf2lamx-1}
\left\{
  \begin{array}{l}
(-1-2\lambda_{(j)}-\kappa_n-\lambda)\lambda_0^{(1)} +\gamma\lambda_0^{(2)}=-N_{2}{\varphi}_2^{(0)}\delta V^{(j)}_0 \\
-\gamma \lambda_0^{(1)}+(-1-2\lambda_{(j)}+\kappa_n-\lambda) 
\lambda_0^{(2)}=-N_{1}{\varphi}_1^{(0)}\delta V^{(j)}_0
  \end{array}
\right.
\end{equation}
We note that in this case this equation in inhomogeneous and has a
non-zero determinant.  The equation for $\zeta_0^{(i)}$ is obtained
from the term of order $1/x^2$ as  
\begin{equation}
  \label{eq:dwfdv0}
\left\{
  \begin{array}{l}
(-\kappa_n-\lambda)\zeta_0^{(1)} +\gamma\zeta_0^{(2)}=0\\
-\gamma \zeta_0^{(1)}+(\kappa_n-\lambda) \zeta_0^{(2)}=0
\end{array}
\right. ,
\end{equation}

\section{Model Potentials}
\label{app:modpot}

One of the models considered here for a screening potential is the 
spherically averaged potential that arises from the charge 
distribution of another electron in state $j$ in the atom:
\begin{eqnarray}
\delta V^{(j)}(x_2) &=& \frac{1}{4\pi}\int d\Omega_2
\int d\bbox{x}_1 \frac{\alpha}{|\bbox{x}_2 - \bbox{x}_1|} 
\left|\phi_j(\bbox{x}_1)\right|^2
\nonumber\\
&=& \alpha\int_0^\infty dx_1 \frac{x_1^2}{ \max(x_2,x_1)}
\nonumber\\&& \qquad\qquad  \times
\left[f_{1,(j)}^2(x_1) + f_{2,(j)}^2(x_1)\right]
\end{eqnarray}
To facilitate numerical integration, this equation is written
as
\begin{eqnarray}
\delta V^{(j)}(x_2) &=& 
\left<\frac{\alpha}{ x}\right> - \alpha\int_0^{x_2} dx_1 x_1^2
\left(\frac{1}{x_1} - \frac{1}{x_2}\right)
\nonumber\\&& \qquad\qquad \times
\left[f_{1,(j)}^2(x_1) + f_{2,(j)}^2(x_1)\right]
\label{eq:mods}
\end{eqnarray}
for $x_2 < x_0$, or as
\begin{eqnarray}
\delta V^{(j)}(x_2) &= &
\frac{\alpha}{ x_2} - \alpha\int_{x_2}^\infty dx_1 x_1^2
\left(\frac{1}{x_2} - \frac{1}{x_1}\right)
\nonumber\\&& \qquad\qquad \times
\left[f_{1,(j)}^2(x_1) + f_{2,(j)}^2(x_1)\right]
\label{eq:modl}
\end{eqnarray}
for $x_2> x_0$, for a suitable value of $x_0$.
The expectation value in (\ref{eq:mods}) is evaluated with the aid of
the identity
\begin{eqnarray}
\left<\frac{\alpha}{ x}\right> = - \frac{\partial E_n}{ \partial Z}
\end{eqnarray}
where $E_n$ is the energy eigenvalue of the screening 
wave function.

The crossover point is taken to be $x_0 = 2/\sqrt{1-E_n^2}$.  The 
integral in (\ref{eq:mods}) is evaluated by 20 point Gauss Legendre 
quadrature with a new integration variable $t$ over the range $(0,1)$ 
defined by $x_1 = x_2t^4$, and the integral in (\ref{eq:modl}) is 
evaluated by 25 Gauss Laguerre quadrature with a new integration 
variable $s$ over the range $(0,\infty)$ where $x_1 = x_2 + 
s/(2\sqrt{1-E_n^2})$.  This prescription gives a precision of better 
than one part in $10^{12}$ for the range $1 \le Z \le 100$, as 
determined by comparing results of the two methods of integration in 
(\ref{eq:mods}) and (\ref{eq:modl}) .

The corresponding derivatives are
\begin{eqnarray}
\lefteqn{\frac{d }{ dx_2} \delta V^{(j)}(x_2) = 
- \frac{\alpha}{ x_2^2}\int_0^{x_2} dx_1 x_1^2} &&\nonumber \\
&&\qquad \qquad \qquad \qquad\times  \left[f_{1,(j)}^2(x_1) + f_{2,(j)}^2(x_1)\right]
\label{eq:dmods}
\end{eqnarray}
for $x_2 < x_0$, or
\begin{eqnarray}
\frac{d}{ dx_2}\delta V^{(j)}(x_2) =
- \frac{\alpha}{ x_2^2} \nonumber \\ \times \left(1 - 
\int_{x_2}^\infty dx_1 x_1^2
\left[f_{1,(j)}^2(x_1) + f_{2,(j)}^2(x_1)\right]
\right)
\label{eq:dmodl}
\end{eqnarray}
for $x_2> x_0$.

The derivatives are calculated with the same integration methods as 
the described above for the function $\delta V^{(j)}(x)$.


A simple additional model potential, useful for testing code, is 
generated by employing an exponential charge distribution, which 
corresponds to the replacement
\begin{eqnarray}
\left[f_{1,(j)}^2(x_1) + f_{2,(j)}^2(x_1)\right]
\rightarrow
4\gamma^3e^{-2\gamma x} \quad ,
\end{eqnarray}
in Eqs.~(\ref{eq:mods}), (\ref{eq:modl}), (\ref{eq:dmods}), and 
(\ref{eq:dmodl}) and leads to the analytic potential
\begin{eqnarray}
\delta V^{\rm exp}(x) = \frac{\alpha}{ x} - \alpha\left(\frac{1}{x}+\gamma\right)
e^{-2\gamma x}
\end{eqnarray}
with
\begin{eqnarray}
\left<\frac{\alpha}{ x}\right> = \alpha\gamma
\end{eqnarray}
and the derivative
\begin{eqnarray}
\frac{d}{ dx}\delta V^{\rm exp}(x) = -\frac{\alpha}{ x^2} 
+ \alpha\left(\frac{1}{x^2}+\frac{2\gamma}{ x} + 2\gamma^2 \right)
e^{-2\gamma x}
\end{eqnarray}

\newpage
\narrowtext
\begin{table}[tp]
        \centering
        \caption{Comparison between the present direct calculation (Dir.) with 
        $\delta V (r)=\alpha/r$ and a calculation using numerical 
        derivative of the one-electron self-energy (Num. Der.). Comparison is done 
        for the scaled function $F(Z\alpha)$ .} 
        \begin{tabular}{cdd}
\multicolumn{3}{c}{$1s$}                                                        \\
\hline                                                  
$Z$     &       \multicolumn{1}{c}{Num. Der.}           &       \multicolumn{1}{c}{Dir.}                \\
20      &       -11.1383        (3)     &       -11.1384        (8)     \\
50      &       -6.35896        (5)     &       -6.3589 (2)     \\
90      &       -6.05030        (3)     &       -6.0502 (5)     \\
\hline                                                  
\multicolumn{3}{c}{$2s$}                                                        \\
\hline                                                  
$Z$     &       \multicolumn{1}{c}{Num. Der.}           &       \multicolumn{1}{c}{Dir.}                \\
20      &       -12.2267        (8)     &       -12.227 (1)     \\
50      &       -8.0243 (1)     &       -8.0243 (3)     \\
90      &       -10.03340       (6)     &       -10.0334        (6)     \\
\hline                                                  
\multicolumn{3}{c}{$2p_{1/2}$}                                                  \\
\hline                                                  
$Z$     &       \multicolumn{1}{c}{Num. Der.}           &       \multicolumn{1}{c}{Dir.}                \\
20      &       0.3190  (8)     &       0.319   (6)     \\
50      &       -0.2444 (1)     &       -0.2445 (8)     \\
90      &       -2.30603        (4)     &       -2.30607        (7)     \\
\hline                                                  
\multicolumn{3}{c}{$2p_{3/2}$}                                                  \\
\hline                                                  
$Z$     &       \multicolumn{1}{c}{Num. Der.}           &       \multicolumn{1}{c}{Dir.}                \\
20      &       -0.606  (1)     &       -0.61   (1)     \\
50      &       -0.9056 (1)     &       -0.906  (2)     \\
90      &       -1.37474        (3)     &       -1.3748 (3)     \\
\end{tabular}
\label{tab:coulcomplh}
\end{table}
\newpage
\widetext
\begin{table}[tbp]
 \centering
 \caption{Self-energy screening function $f(Z\alpha)$ for $1s$, $2s$, 
 $2p_{1/2}$ and $2p_{3/2}$ electrons}
 \begin{tabular}{c|dddd}
        &       \multicolumn{4}{c}{$F(Z\alpha)$ for $1s$ screened by}                                                                                   \\
$Z$     &       $1s$            &       $2s$            &       $2p_{1/2}$              &       $2p_{3/2}$              \\
\hline                                                                                                  
5       &       -5.171  (5)     &       -0.491  (2)     &       -0.243  (2)     &       -0.242  (2)     \\
10      &       -3.882  (1)     &       -0.3681 (4)     &       -0.1832 (5)     &       -0.1809 (5)     \\
18      &       -2.901  (1)     &       -0.27670        (1)     &       -0.14115        (1)     &       -0.13555        (1)     \\
20      &       -2.7386 (3)     &       -0.26184        (10)    &       -0.1345 (1)     &       -0.1279 (1)     \\
24      &       -2.4726 (3)     &       -0.23772        (7)     &       -0.12394        (8)     &       -0.11536        (8)     \\
30      &       -2.1726 (1)     &       -0.21111        (4)     &       -0.11286        (5)     &       -0.10092        (5)     \\
32      &       -2.0917 (1)     &       -0.20410        (4)     &       -0.11010        (5)     &       -0.09696        (5)     \\
40      &       -1.83507        (8)     &       -0.18267        (3)     &       -0.10245        (3)     &       -0.08403        (3)     \\
44      &       -1.73770        (6)     &       -0.17506        (2)     &       -0.10028        (3)     &       -0.07891        (2)     \\
50      &       -1.62045        (5)     &       -0.16663        (2)     &       -0.09870        (2)     &       -0.07243        (2)     \\
54      &       -1.55825        (4)     &       -0.16272        (1)     &       -0.09867        (2)     &       -0.06875        (2)     \\
60      &       -1.48502        (4)     &       -0.15912        (1)     &       -0.10011        (1)     &       -0.06402        (1)     \\
66      &       -1.43296        (4)     &       -0.15809        (1)     &       -0.10334        (1)     &       -0.06003        (1)     \\
70      &       -1.40906        (3)     &       -0.158821       (9)     &       -0.10661        (1)     &       -0.057727       (10)    \\
74      &       -1.3935 (2)     &       -0.160737       (9)     &       -0.11085        (1)     &       -0.055657       (9)     \\
80      &       -1.38620        (2)     &       -0.166065       (8)     &       -0.119409       (9)     &       -0.052952       (8)     \\
83      &       -1.39010        (7)     &       -0.16998        (1)     &       -0.124865       (10)    &       -0.051754       (7)     \\
90      &       -1.4214 (2)     &       -0.18313        (4)     &       -0.14160        (4)     &       -0.04933        (2)     \\
92      &       -1.43690        (2)     &       -0.188136       (6)     &       -0.147674       (8)     &       -0.048726       (6)     \\
\hline                                                                                                  
        &       \multicolumn{4}{c}{$F(Z\alpha)$ for $2s$ screened by}                                                                                   \\
        &       $1s$            &       $2s$            &       $2p_{1/2}$              &       $2p_{3/2}$              \\
\hline                                                                                                  
10      &       -7.951  (9)     &       -3.162  (6)     &       -4.180  (6)     &       -4.171  (6)     \\
20      &       -5.770  (1)     &       -2.2806 (8)     &       -3.0141 (9)     &       -2.9876 (9)     \\
30      &       -4.740  (1)     &       -1.8624 (4)     &       -2.4615 (5)     &       -2.4120 (5)     \\
40      &       -4.1691 (4)     &       -1.6260 (2)     &       -2.1503 (2)     &       -2.0717 (2)     \\
50      &       -3.8535 (3)     &       -1.4896 (1)     &       -1.9720 (1)     &       -1.8563 (1)     \\
60      &       -3.7137 (3)     &       -1.4208 (1)     &       -1.88388        (9)     &       -1.71942        (8)     \\
70      &       -3.7207 (2)     &       -1.4072 (2)     &       -1.86978        (8)     &       -1.63864        (9)     \\
80      &       -3.87690        (9)     &       -1.44860        (6)     &       -1.92989        (6)     &       -1.60346        (6)     \\
90      &       -4.2184 (2)     &       -1.55689        (8)     &       -2.08125        (9)     &       -1.61098        (8)     \\
92      &       -4.3155 (2)     &       -1.58889        (8)     &       -2.12572        (10)    &       -1.61786        (8)     \\
\hline                                                                                                  
        &       \multicolumn{4}{c}{$F(Z\alpha)$ for $2p_{1/2}$ screened by}                                                                                     \\
        &       $1s$            &       $2s$            &       $2p_{1/2}$              &       $2p_{3/2}$              \\
\hline                                                                                                  
10      &       0.10    (6)     &       -0.03   (2)     &       -0.06   (2)     &       -0.06   (2)     \\
20      &       0.064   (5)     &       -0.027  (2)     &       -0.049  (2)     &       -0.049  (2)     \\
30      &       -0.012  (1)     &       -0.0373 (9)     &       -0.0620 (6)     &       -0.0611 (6)     \\
40      &       -0.119  (5)     &       -0.056  (3)     &       -0.088  (4)     &       -0.084  (4)     \\
50      &       -0.2566 (5)     &       -0.0817 (3)     &       -0.1250 (2)     &       -0.1158 (2)     \\
60      &       -0.4316 (2)     &       -0.1164 (2)     &       -0.1757 (1)     &       -0.1547 (1)     \\
70      &       -0.6607 (9)     &       -0.1636 (7)     &       -0.245  (7)     &       -0.202  (3)     \\
80      &       -0.9751 (5)     &       -0.2300 (6)     &       -0.3446 (9)     &       -0.2604 (5)     \\
90      &       -1.4367 (1)     &       -0.33002        (7)     &       -0.49508        (7)     &       -0.33416        (6)     \\
92      &       -1.5564 (2)     &       -0.3563 (2)     &       -0.5348 (1)     &       -0.35143        (9)     \\
\hline                                                                                                  
        &       \multicolumn{4}{c}{$F(Z\alpha)$ for $2p_{3/2}$ screened by}                                                                                     \\
        &       $1s$            &       $2s$            &       $2p_{1/2}$              &       $2p_{3/2}$              \\
\hline                                                                                                  
10      &       -0.7    (2)     &       -0.21   (9)     &       -0.3    (1)     &       -0.3    (1)     \\
20      &       -0.663  (3)     &       -0.185  (3)     &       -0.285  (3)     &       -0.281  (3)     \\
30      &       -0.691  (4)     &       -0.189  (2)     &       -0.287  (2)     &       -0.279  (2)     \\
40      &       -0.737  (1)     &       -0.1992 (3)     &       -0.3004 (4)     &       -0.2849 (4)     \\
50      &       -0.794  (1)     &       -0.2154 (4)     &       -0.3223 (4)     &       -0.2962 (4)     \\
60      &       -0.8599 (4)     &       -0.2369 (2)     &       -0.3524 (2)     &       -0.3111 (2)     \\
70      &       -0.9330 (3)     &       -0.2642 (2)     &       -0.3912 (5)     &       -0.3286 (3)     \\
80      &       -1.0123 (3)     &       -0.299  (1)     &       -0.440  (1)     &       -0.3486 (8)     \\
90      &       -1.0972 (2)     &       -0.3419 (2)     &       -0.5022 (1)     &       -0.3704 (1)     \\
92      &       -1.114  (9)     &       -0.351  (6)     &       -0.516  (6)     &       -0.374  (6)     \\
\end{tabular} 
\label{tbl:totcont}
\end{table}
\begin{table}[tbp]
        \centering
        \caption{Comparison between $F(Z\alpha)$ using the
        partial-wave renormalization method \protect \cite{pssl96} and
        this work. More accurate numerical values from Ref.~\protect
        \cite{sun98} are used. $R_{\textrm{rms}}$ is the nuclear
        mean-spherical charge radius used (in Fm). Since no error
        estimate is provided with Ref.~\protect \cite{sun98}, in
        which $2F(Z\alpha)$ is tabulated we use an error of 1 on the
        last displayed figure.}
        \begin{tabular}{cdddd}
$Z$&  \multicolumn{1}{c}{$R_{\textrm{rms}}$} &  \multicolumn{1}{c}{Ref.~\cite{sun98}}  & \multicolumn{1}{c}{This work}  & \multicolumn{1}{c}{Diff.}  \\
\tableline                                                                                              
5       &               &       -5.1745 (5)     &       -5.171  (5)     &       0.004   (5)     \\
10      &               &       -3.8820 (5)     &       -3.882  (1)     &       0.000   (1)     \\
18      &       3.423   &       -2.89995        (5)     &       -2.901  (1)     &       -0.001  (1)     \\
24      &       3.643   &       -2.47240        (5)     &       -2.4726 (3)     &       -0.0002 (3)     \\
32      &       4.07    &       -2.09145        (5)     &       -2.0917 (1)     &       -0.0002 (1)     \\
44      &       4.480   &       -1.73720        (5)     &       -1.73770        (6)     &       -0.00050        (8)     \\
54      &       4.78    &       -1.55720        (5)     &       -1.55825        (4)     &       -0.00105        (7)     \\
66      &       5.21    &       -1.43035        (5)     &       -1.43296        (4)     &       -0.00261        (6)     \\
74      &       5.37    &       -1.38890        (5)     &       -1.3935 (2)     &       -0.0046 (3)     \\
83      &       5.519   &       -1.38080        (5)     &       -1.39010        (7)     &       -0.00930        (8)     \\
92      &       5.860   &       -1.41630        (5)     &       -1.43690        (2)     &       -0.02060        (5)     \\
\end{tabular} 
\label{tbl:compsun}
\end{table}
\begin{table}[tbp]
        \squeezetable
        \centering
        \caption{Comparison of the present result for lithium-like ions with earlier 
        work (eV). Results from Refs. \protect 
        \cite{yab98,yabss98,yab99,bcs97,blu92,blu93b,cjs93,plss93} all include finite nuclear size 
        correction, not included in the present work. We use results without exchange. The difference with Ref.~\protect\cite{iam91} 
        on the $2p_{1/2}$ screening is due to a programming error in the code used in 1991.}
        \begin{tabular}{ccdddddddd}
Orbital &       screened by     &       \multicolumn{1}{c}{$\Delta E$}          &       
\multicolumn{1}{c}{Ref. \cite{iam91}}   &       \multicolumn{1}{c}{Diff.}       &       \multicolumn{1}{c}{Refs. \cite{blu92,blu93b}}           &       \multicolumn{1}{c}{Ref. \cite{bcs97}}   &       \multicolumn{1}{c}{Ref. \cite{yabss98}} &       \multicolumn{1}{c}{Ref. \cite{cjs93}}           &       \multicolumn{1}{c}{Ref. \cite{plss93}}          \\
\tableline
$1s$    &       $1s$    &       -3.76607        (5)     &       -3.79   &       0.02    &                       &               &               &                       &                       \\
        &       $2s$    &       -0.49310        (2)     &       -0.48   &       -0.01   &                       &               &               &                       &                       \\
        &       $2p_{1/2}$      &       -0.38705        (2)     &       -0.33   &       -0.06   &                       &               &               &                       &                       \\
        &       $2p_{3/2}$      &       -0.12771        (2)     &               &               &                       &               &               &                       &                       \\
$2s$    &       $1s$    &       -1.41385        (8)     &       -1.44   &       0.03    &       -1.375  (5)     &       -1.389  &       -1.39   &       -1.375  (30)    &       -1.385  (15)    \\
$2p_{1/2}$      &       $1s$    &       -0.50991        (5)     &       -0.32   &       -0.19   &       -0.485  (5)     &       -0.506  &       -0.505  &       -0.475  (30)    &       -0.495  (15)    \\
$2p_{3/2}$      &       $1s$    &       -0.365  (3)     &               &               &                       &       -0.356  &               &                       &                       \\
\tableline                                                                                                                                                                                      
$1s^2\,2s$      &               &       -11.3460        (1)     &       -11.42  &       0.08    &                       &               &               &                       &                       \\
$1s^2\,2p_{1/2}$        &               &       -9.32606        (8)     &       -8.88   &       -0.45   &                       &               &               &                       &                       \\
$1s^2\,2p_{3/2}$        &               &       -8.518  (4)     &               &               &                       &               &               &                       &                       \\
\tableline                                                                                                                                                                                      
\multicolumn{3}{c}{Transitions (Valence+Core)}                                          &               &               &                       &               &               &                       &                       \\
$2p_{1/2}\to 2s$        &               &       2.0200  (2)     &       2.55    &       -0.53   &       2.01    (1)     &               &               &                       &                       \\
$2p_{3/2}\to 2s$        &               &       2.828   (4)     &               &               &                       &               &               &                       &                       \\
\tableline                                                                                                                                                                                      
\multicolumn{3}{c}{Transitions (Valence)}                                               &               &               &                       &               &               &                       &                       \\
$2p_{1/2}\to 2s$        &               &       1.8079  (1)     &               &               &       1.78    (1)     &       1.766   &       1.77    &       1.80    (6)     &       1.78    (3)     \\
$2p_{3/2}\to 2s$        &               &       2.098   (4)     &               &               &                       &       2.065   &               &                       &                       \\
\end{tabular}
\label{tbl:lilikeu}
\end{table}

\widetext

\newpage
\newpage
\newpage
\newpage


\end{document}